\documentclass[amsmath,amssymb,aps,prapplied,reprint,showpacs,twocolumn,floats,superscriptaddress]{revtex4-1}
\usepackage{graphicx}
\usepackage{amssymb}
\usepackage{amsmath}
\usepackage{hyperref}
\usepackage{footnote}
\usepackage{color}
\usepackage{multirow}
\usepackage{enumerate}
\usepackage{soul}
\usepackage[utf8]{inputenc}

\begin{document}
\title{Perturbative diagonalization for time-dependent strong interactions}
\author{Z.~Xiao}
\email{zhihao\_xiao@uml.edu}
\affiliation{Department of Physics and Applied Physics, University of Massachusetts, Lowell, MA 01854, USA}
\author{E.~Doucet}
\affiliation{Department of Physics and Applied Physics, University of Massachusetts, Lowell, MA 01854, USA}
\author{T.~Noh}
\affiliation{Department of Physics and Applied Physics, University of Massachusetts, Lowell, MA 01854, USA}
\affiliation{Associate of the National Institute of Standards and Technology, Boulder, Colorado 80305, USA}
\author{L.~Ranzani}
\affiliation{Quantum Engineering and Computing, Raytheon BBN Technologies, Cambridge, MA 02138, USA}
\author{R.~W.~Simmonds}
\affiliation{National Institute of Standards and Technology, 325 Broadway, Boulder, CO 80305, USA}
\author{L.~C.~G.~Govia}
\affiliation{Quantum Engineering and Computing, Raytheon BBN Technologies, Cambridge, MA 02138, USA}
\author{A.~Kamal}
\email{archana\_kamal@uml.edu}
\affiliation{Department of Physics and Applied Physics, University of Massachusetts, Lowell, MA 01854, USA}
\begin{abstract}
We present a systematic method to implement a perturbative Hamiltonian diagonalization based on the time-dependent Schrieffer-Wolff transformation. Applying our method to strong parametric interactions we show how, even in the dispersive regime, full Rabi model physics is essential to describe the dressed spectrum. Our results unveil several qualitatively new results including realization of large energy-level shifts, tunable in magnitude and sign with the frequency and amplitude of the pump mediating the parametric interaction. Crucially Bloch-Siegert shifts, typically thought to be important only in the ultra-strong or deep-strong coupling regimes, can be rendered large even for weak dispersive interactions to realize points of exact cancellation of dressed shifts (`blind spots') at specific pump frequencies. The framework developed here highlights the rich physics accessible with time-dependent interactions and serves to significantly expand the functionalities for control and readout of strongly-interacting quantum systems.
\end{abstract}
\maketitle
%
\section{Introduction}
%
Time-dependent interactions provide a powerful paradigm for the study of out-of-equilibrium quantum matter. Though originally motivated by the interest in the exploration of new phases in many-body quantum systems, recent developments in this domain have largely been triggered by studies of low-dimensional platforms, such as trapped ions, cavity- and circuit-QED, where a precise control of quantum dynamics is essential for realizing high-fidelity quantum information processing. However, the theoretical framework to describe the dynamics in the presence of such time-dependent interactions remains rudimentary, especially in the strong-coupling regime relevant to most applications. 

One approach to studying strongly coupled systems is to derive a low-energy effective Hamiltonian by using a unitary transformation that decouples the high-frequency (`fast') subspace from the low-energy (`slow') subspace. A standard and widely used method to implement this is the Schrieffer-Wolff transformation (SWT) \cite{Bravyi2011} which develops the diagonalized Hamiltonian as a perturbation series, whose radius of convergence is dictated by the ratio of interaction strength ($g$) and the gap between the low- and high-energy subspaces ($\Delta$). In contrast to conventional Dyson series expansion, SWT method gives direct access to effective Hamiltonian at each order in perturbation series; inferring the effective Hamiltonian from time-ordered matrix exponential describing the propagator is typically non-trivial, especially because truncating the Dyson series does not preserve unitarity, and is shown to fail for degenerate ground states \cite{Brouder2008}

In this work, we present a generalization of the SWT for time-dependent strong interactions. While this exercise has been attempted before in a few sporadic examples \cite{Bukov2016, Petrescu2020, Malekakhlagh2020, Sentef2020}, our approach clarifies how the inertial term generated due to the time-dependent generator of the SWT (equivalent to an extra dynamical rotation of the eigenbasis) can be accounted for at successive orders in a perturbation series; this is especially crucial since its inclusion magnifies the corrections arising from counter-rotating terms that are typically neglected in the dispersive regime ($g/\Delta \ll 1$).
\par
We emphasize that the  approach we describe is very general, and can be applied to a wide family of time-dependent Hamiltonians. Here we demonstrate it for the two archetypal examples of strong light-matter interaction: the Rabi model and a Kerr oscillator coupled to a linear oscillator by a parametrically-modulated interaction. Such periodic driving is of interest in a myriad of applications, such as the realization of synthetic gauge fields in optical lattices \cite{Barbiero2019}, simulation of topological \cite{Roushan2016,Peano2016} and dynamically localized phases \cite{Leuch2016,Mathey2019}, implementation of fast entangling gates \cite{Reagor2018}, tunable qubit readout \cite{Noh2021}, and state stabilization \cite{Lu2017,Doucet2020} and transfer \cite{Sirois2015,Li2018}. 
\par
The paper is organized as follows: We begin with a general description of the diagonalization procedure employing a series of sequential SWT generators and present the equation of motion for constructing time-dependent SWT generator at a given order in perturbation in Sec.~\ref{sec:SWT}. Then, in Sec.~\ref{sec:Rabi}, we apply this method to time-dependent Rabi Hamiltonian and explicitly derive the condition for validity of dispersive approximation in the presence of a parametrically-mediated interaction in this system. We also derive analytical expressions for leading-order dispersive shifts, and comment on their unique features distinct from the `usual' (but time-dependent) Jaynes-Cummings case. We then discuss a generalization of the parametric QED system to include multi-level effects by considering the case of a Kerr oscillator (a.k.a. transmon) parametrically coupled to a linear oscillator and identify a `parametric straddling' regime by studying the induced frequency shifts as a function of pump frequency in Sec.~\ref{sec:Kerr}. Sec.~\ref{sec:Purcell} is devoted to the discussion of induced dissipation in the systems studied in Secs.~\ref{sec:Rabi} and \ref{sec:Kerr}. Finally, we conclude with a summary of results and present potential directions along which the current analysis can be extended in Sec.~\ref{sec:conclusion}. Appendices \ref{app:SWT}-\ref{app:purcell} include additional mathematical details and proofs.
%
%
\section{Time-dependent Schrieffer-Wolff Transformation (SWT)}
\label{sec:SWT}
%
%
For a system described by a time-dependent Hamiltonian $H(t) = H_{0} + \lambda V(t)$, an effective diagonal Hamiltonian at a given order $M$ can be obtained by performing a sequence of ${M-1}$ successive time-dependent Schrieffer-Wolff transformations (SWTs). To this end the aim is to thus construct a unitary operator $\hat{U}(t)$
perturbatively, as a series of successive time-dependent rotations
\begin{eqnarray}
 \hat{U}(t) = \prod_{j=1}^{M-1} \exp[\hat{S}^{(j)}(t)],
 \label{eqn:SWTket}
\end{eqnarray}
where $\hat{S}^{(n)}(t) \sim \mathcal{O}(\lambda^{n})$ are the anti-Hermitian SWT generators. Under the action of $U(t)$ the lab frame Hamiltonian $\hat{H}(t)\equiv\hat{H}^{(1)}(t)$ is transformed into $\hat{H}^{(M)}(t)$,
\begin{eqnarray}
\hat{H}^{(M)}(t)=&\hat{H}_0+\hat{V}^{(M)}(t).
\label{Eq:Condition}
\end{eqnarray}
Then, the effective diagonalized (`low-energy') Hamiltonian to order $\lambda^{M}$ is given as,
\begin{eqnarray}
    \hat{H}_{\rm eff}^{(M)}(t) \equiv \hat{H}_{0}+\mathcal{P}_{0}\bullet\hat{V}^{(M)}(t) = \hat{H}_{0} + \hat{V}^{(M)}_{\text{D}}(t),
\end{eqnarray}
where
\begin{eqnarray}
    \mathcal{P}_{0}\bullet\hat{\Theta} = \sum_k|\psi_k\rangle\langle\psi_k|\hat{\Theta}|\psi_k\rangle\langle\psi_k|,
\end{eqnarray}
is the pinching channel in the eigenbasis $\{|\psi_k\rangle\}$ of $\hat{H}_{0}$, which defines the `low-energy' subspace for the purpose of diagonalization.
\par
The off-diagonal part of the interaction Hamiltonian contains terms of $\mathcal{O}(\lambda^M)$ or higher, i.e. 
\begin{eqnarray}
\hat{V}_{\text{OD}}^{(M)}(t)\equiv \mathcal{Q}_{0} \bullet \hat{V}^{(M)}(t)=\sum\limits_{m=M}^{\infty}\lambda^{m}\hat{V}_{\text{OD},m}^{(M)}(t),
\label{Eq:VOD}
\end{eqnarray}
where
\begin{eqnarray}
     \mathcal{Q}_{0} \bullet \hat{\Theta} = \sum_{j\neq k}|\psi_j\rangle\langle\psi_j|\hat{\Theta}|\psi_k\rangle\langle\psi_k|.
\end{eqnarray}
projects onto the off-diagonal elements in the eigenbasis of $\hat{H}_0$. Such a construction is achievable by finding the operators $\hat{S}^{(j)}(t)$ using the following theorem. 
\\
\par
\noindent{\bf Theorem}: At each order $M$ in perturbation series, if $\exists \; \hat{S}^{(M)}(t)$ ($M \geq 1$) satisfying the differential equation 
\begin{equation}
\label{eqn:S_Generator}
i\frac{\partial\hat{S}^{(M)}(t)}{\partial t}+[\hat{S}^{(M)}(t),\hat{H}_0]+\lambda^M\hat{V}_{\text{OD},M}^{(M)}(t)= 0,
\end{equation} 
then $\hat{U}_{M}(t)=\exp[-\hat{S}^{(M)}(t)]$ can be used to eliminate off-diagonal terms in $\hat{H}^{(M)}(t)$ up to $\mathcal{O}(\lambda^M)$, s.t. 
\begin{eqnarray}
{\hat{H}^{(M+1)}(t) =  \hat{H}_{\rm eff}^{(M)}(t) + \mathcal{O}(\lambda^{M+1})},
\end{eqnarray}
where $\hat{H}_{\rm eff}^{(M)}(t) \equiv \hat{H}_{0}+\mathcal{P}_{0}\bullet\hat{V}^{(M)}(t) = \hat{H}_{0} + \hat{V}^{(M)}_{\text{D}}(t)$. Here $\hat{V}_{\text{OD},M}^{(M)}(t)$ denotes the leading off-diagonal term at $\mathcal{O}(\lambda^{M})$ in the perturbation series, and includes the terms generated by the action of lower-order time-dependent transformations on $V(t)$.
Note that with $\hat{H}_{\rm eff}^{(M)}(t)$ being diagonal, the off-diagonal part and the remaining diagonal part of $\hat{H}^{(M+1)}(t)$ are both $\mathcal{O}(\lambda^{M+1})$. In this way, $M$-th order of the off-diagonal part of $\hat{H}^{(M)}(t)$ is eliminated as it is transformed into $\hat{H}^{(M+1)}(t)$, with
\begin{equation}
\hat{H}^{(M+1)}(t)=\hat{U}_{M}^{\dagger}(t)\hat{H}^{(M)}(t)\hat{U}_{M}(t)-i\hat{U}_{M}^{\dagger}(t)\frac{\partial\hat{U}_{M}(t)}{\partial t}.
\end{equation} 
See appendix~\ref{app:SWT} for a detailed proof and a specific example. 
\par
It is worthwhile to note that typical methods of perturbative diagonalization develop corrections at different orders by expanding a single generator as a series in $\lambda$, i.e. $\hat{S}^{(M)}(t) = \exp[\sum_{n=1}^{M-1} \lambda^{n}G^{(n)}(t)]$, which leads to a set of coupled differential equations for generators $G^{(n)}(t)$ ($n\geq 3 $) for explicitly time-dependent interactions \cite{Malekakhlagh2020}. On the other hand, the construction presented here allows cancellation of the derivative term at each order in $\lambda$ {\it independently} since the orders are tracked by updating the off-diagonal contribution at each step, which greatly simplifies the equations of motion for higher-order generators [see appendix~\ref{app:SWT}]. 
%
%
\section{Parametric Light-Matter Interactions}
\label{sec:Rabi}
\subsection{Rabi Hamiltonian}
\begin{figure}[t!]
	\centering
	\includegraphics[width=0.95\columnwidth]{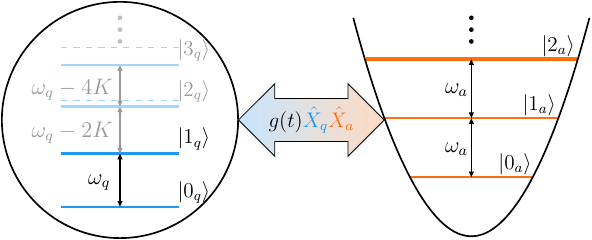}
	\caption{
	Schematic representation of a bipartite system with strong time-dependent coupling. For the qubit case we confine ourselves to the lowest two levels, while for a multi-level generalization, we consider the energy level ladder of a Kerr oscillator with anharmonicity $K$.
	}
    \label{Fig:SchematicSystemQR}%
\vspace{-5pt}
\end{figure}

%
We now apply the sequential SWT construction described in the previous section to a time-dependent Rabi Hamiltonian describing a qubit-resonator system (Fig.~\ref{Fig:SchematicSystemQR}). The system Hamiltonian with time-dependent transverse coupling is given by ($\hbar= 1$)
\begin{equation}\begin{split}
\hat{H}(t) =-\frac{\omega_q}{2}\hat{\sigma}_q^z+\omega_a\left(\hat{a}^\dagger\hat{a}+\frac{1}{2}\right) + g(t)\hat{\sigma}_q^x(\hat{a}^\dagger+\hat{a})
\label{Eq:RabiH}
\end{split}\end{equation}
For a block off-diagonal interaction of the form in Eq.~(\ref{Eq:RabiH}), the first time-dependent unitary rotation in our SWT series is described by the generator,
\begin{equation}\label{eqn:SWT_Opertor}
{\hat{S}^{(1)}(t)={\xi}^{(1)}_{+}(t)\hat{\sigma}_q^{+}\hat{a}^\dagger- {\xi}^{(1)}_{-}(t)\hat{\sigma}_q^{-}\hat{a}^\dagger-\text{h.c}},
\end{equation}
where the time-dependence of $V^{(1)}(t)\equiv g(t)\hat{\sigma}_q^x(\hat{a}^\dagger+\hat{a})$ is ensconced in the solution of the coefficients ${\xi}^{(1)}_{\pm}(t)$. These can be evaluated using Eq.~(\ref{eqn:S_Generator}) for $M=1$, which leads to the following set of linear differential equations,
\begin{equation}\begin{split}\label{eqn:beta_condition}
\pm i\dot{\xi}^{(1)}_{\pm}(t)-(\omega_q\pm\omega_a){\xi}^{(1)}_{\pm}(t)+ g(t)=0.
\end{split}\end{equation}
\par
The time-dependent Hamiltonian correct to second-order in $g_{p}$, $\hat{H}^{(2)}(t)$ in the interaction picture, defined w.r.t the time-independent $\hat{H}_{0}$, can then be obtained as
\begin{equation}\begin{split} \label{eqn:H_eff_full}
\hat{H}^{(2)}(t)&=\frac{1}{2}[\hat{S}^{(1)}(t),\hat{V}^{(1)}(t)] +\mathcal{O}(g_{p}^{3})\\
&= -g(t)\left[\text{Re}\{{\xi}^{(1)}_{-}(t)\}+\text{Re}\{{\xi}^{(1)}_{+}(t)\}\right]\hat{\sigma}_q^z\left(\hat{a}^\dagger\hat{a}+\frac{1}{2}\right)\\
&\quad-\frac{g(t)}{2}\left[\left({\xi}^{(1)*}_{-}(t)+{\xi}^{(1)*}_{+}(t)\right)e^{-2 i \omega_at}\hat{a}^2 \hat{\sigma}_q^z+\text{h.c.}\right]\\
&\quad +\frac{g(t)}{2}\left(\text{Re}\{{\xi}^{(1)}_{-}(t)\}-\text{Re}\{{\xi}^{(1)}_{+}(t)\}\right) + \mathcal{O}(g^{3}(t)).
\end{split}\end{equation}
Assuming the interaction time-dependence to be sinusoidal and decomposing it into distinct frequency components:
\begin{equation}\label{eqn:g_general}
g(t)=\sum_{p=1}^{K}[g_p \exp(-i\omega_pt)+g_p^* \exp(+i\omega_pt)],
\end{equation}
this leads to the solution for Eq.~(\ref{eqn:beta_condition}) as
\begin{equation}\begin{split}
\label{eqn:gamma_general}
\xi^{(1)}_{\pm}(t)&=\sum_{p=1}^{K}\left[\frac{g_p}{\omega_q\pm\omega_a\mp\omega_p}\exp(-i\omega_pt)\right.\\
&\left. \hspace{2cm}+\frac{g_p^*}{\omega_q\pm\omega_a\pm\omega_p}\exp(+i\omega_pt)\right].
\end{split}\end{equation}
For the special case of monochromatic parametric driving ($K=1$), Eq.~(\ref{eqn:H_eff_full}) simplifies to
\begin{widetext}
\begin{equation}\begin{split}\label{eqn:H_eff_int_SingleFrequency}
\hat{H}^{(2)}(t)&=-2|g_p|^2\sum_{\pm}^{}\left[\left(\frac{1}{\omega_{\pm}+\omega_p}+\frac{1}{\omega_{\pm}-\omega_p}\right)\left(\hat{a}^\dagger\hat{a}+\frac{1}{2}\right)\hat{\sigma}_q^z+\frac{1}{2}\left(\frac{1}{\omega_{\pm}+\omega_p}+\frac{1}{\omega_{\pm}-\omega_p}\right)\right]\cos^2(\omega_pt+\phi_p)\\
& \qquad -\left[\left\{e^{-2i\omega_a t}|g_p|^2\sum_{\pm}^{}\left(\frac{1}{\omega_{\pm}+\omega_p}+\frac{1}{\omega_{\pm}-\omega_p}\right)+ e^{-2i(\omega_p+\omega_a) t}g_p^2\left(\frac{1}{\omega_{-}-\omega_p}+\frac{1}{{\omega_{+}+\omega_p}}\right)\right.\right.\\
&\qquad\qquad\left.\left.+ \, e^{+2i(\omega_p-\omega_a) t}\left(g_p^*\right)^2\left(\frac{1}{\omega_{-}+\omega_p}+\frac{1}{\omega_{+}-\omega_p}\right)\right\}\hat{a}^2 +\text{h.c.}\right]\frac{\hat{\sigma}_q^z}{2}+\mathcal{O}(g_p^{3}),
\end{split}\end{equation}
\end{widetext}
where $\phi_p=\arg \left(g_p\right)$ and $\omega_{\pm}=\omega_q\pm\omega_a$, with $+/-$ denoting the sum and difference of the qubit and the resonator frequencies respectively. The squeezing terms in Eq.~(\ref{eqn:H_eff_int_SingleFrequency}) are fast-rotating for pump frequencies near $\omega_{\pm}$ and their effect can hence be neglected as their time-average vanishes. On the other hand, when $\omega_p\sim\pm\omega_a$ and the squeezing terms are nearly resonant, the effective detuning is comparable to $\omega_{q}$ which is in GHz; this makes the effect of these terms negligible since $|g_p|/\omega_q \ll 1$. By a similar argument, cubic contributions towards $\chi_{p}$ from $g_p^3$ can be neglected under the RWA. This order necessarily involves terms of the form $[\hat{S}^{(1)},[\hat{S}^{(1)},\hat{V}^{(1)}]]$ 
, and each term of this form contains an asymmetric number of raising and lowering operators ${(\hat{a}^\dagger, \hat{a})}$ such that it can only be rendered resonant for large detunings from parametric resonance. 
\par
Consequently, the diagonal part of the Hamiltonian in Eq.~(\ref{eqn:H_eff_int_SingleFrequency}) to order $\mathcal{O}(g_p^{2})$, $\hat{H}_\text{eff}^{(2)}(t)=\mathcal{P}_{0}\bullet\frac{1}{2}[\hat{S}^{(1)}(t),\hat{V}^{(1)}(t)]$ simplifies to its time-averaged form as,
\begin{equation}\begin{split}\label{eqn:H_eff_int_SingleFrequency_RWA}
\hat{H}_\text{eff}^{(2)}=\chi_p^{(2)}\left(\hat{a}^\dagger\hat{a}+\frac{1}{2}\right)\hat{\sigma}_q^z+\Omega_{p}^{(2)},
\end{split}\end{equation}
with
\begin{equation}\begin{split}\label{eqn:QubitDispersiveShift}
\chi_p^{(2)}=-2|g_p|^2 \left(\frac{\omega_{-}}{\omega_{-}^2-\omega_p^2}+\frac{\omega_{+}}{\omega_{+}^2-\omega_p^2}\right)
\end{split}\end{equation} 
denoting the leading-order dispersive shift and
\begin{equation}\begin{split}
\Omega_{p}^{(2)}=|g_p|^2\left(\frac{\omega_{-}}{\omega_{-}^2-\omega_p^2}-\frac{\omega_{+}}{\omega_{+}^2-\omega_p^2}\right).
\end{split}\end{equation}
being the overall energy shift.
\par
\begin{figure}[t!]
\centering
\includegraphics[width=\columnwidth]{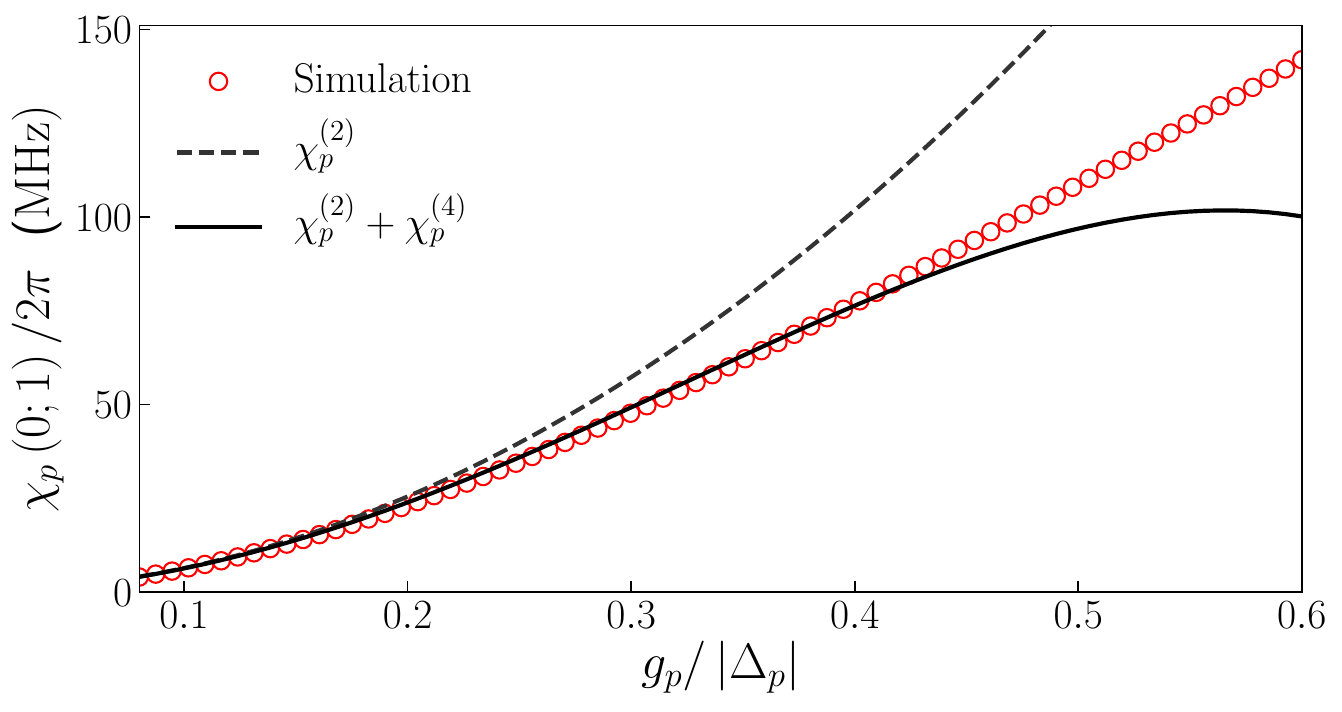}%
\vspace{-5pt}
 \caption{Comparison of analytical calculations for the linear dispersive shift, obtained from $\chi_{p}\hat{a}^{\dagger}\hat{a}\hat{\sigma}_{q}^{z}$ term in $\hat{H}_\text{eff}^{(2)}(t)$ and $\hat{H}_\text{eff}^{(4)}(t)$ respectively, with exact results obtained from numerically simulated resonator spectrum. All simulations were performed with $\omega_q=2\pi\times5$ GHz, $\omega_a=2\pi\times3$ GHz, ${\omega_p=2\pi\times1.5}$~GHz, and $\kappa=2\pi\times1.5$ MHz.}
\label{Fig:chi4vschi2}%
\vspace{-5pt}
\end{figure}

To eliminate $\mathcal{O}(g_p^{2})$ off-diagonal terms in Eq.~(\ref{eqn:H_eff_int_SingleFrequency}), we develop the next-order SWT generator as ${\hat{S}^{(2)}(t)={\xi}^{(2)}(t)\hat{\sigma}_q^{z}(\hat{a}^\dagger)^{2}-\text{h.c.}}$, where ${\xi}^{(2)}(t)$ is again evaluated using Eq.~(\ref{eqn:S_Generator}) for ${M=2}$. The resultant effective Hamiltonians are [see appendix~\ref{app:higherorder} for detailed expressions of higher-order shifts]:
\begin{eqnarray}\label{Eq:H2H4}
  \hat{H}_\text{eff}^{(3)}(t) &=& 0\\
  \hat{H}_\text{eff}^{(4)}(t) &=& \mathcal{P}_{0}\bullet\frac{1}{8}\left[\hat{S}^{(1)}(t),\big[\hat{S}^{(1)}(t),[\hat{S}^{(1)}(t),\hat{V}^{(1)}(t)]\big]\right] \nonumber\\
  & & + \mathcal{P}_{0}\bullet\frac{1}{2}\left[\hat{S}^{(2)}(t), \mathcal{Q}_{0}\bullet \frac{1}{2}\Big[\hat{S}^{(1)}(t),\hat{V}^{(1)}(t)\Big]\right],\nonumber
\end{eqnarray}
Figure~\ref{Fig:chi4vschi2} shows a comparison of dressed shifts predicted from  Eq.~(\ref{Eq:H2H4}) with exact shifts estimated from a numerical simulation of the parametric Rabi Hamiltonian. Note that for this specific case, the presence of the inertial term significantly modifies the magnitude of $|{\xi}^{(1)}_{\pm}|$, and hence the condition for validity of the perturbative expansion $|{\xi}^{(1)}_{\pm}| \ll1$ leads to
\begin{equation}
|g_{p}|\ll\Delta_{p};\; \Delta_{p} = {\rm min}\{|\omega_{\pm} \pm \omega_{p}|\}.
\label{eqn:validdisp}
\end{equation}
This implies that the interaction strength needs to be small compared to the detunings measured in a rotating frame defined w.r.t. the pump frequency $\omega_{p}$. This presents an opportunity to realize large dressed shifts, in excess of 10 MHz with $|g_{p}|$ of only a few tens of MHz, as evident from the results presented in Fig.~\ref{Fig:chi4vschi2}. For comparison, achieving a similar value of the shift with static couplings would require an interaction strength of magnitude comprable to the qubit-resonator detuning (i.e. $|g_{p}| \sim \omega_{-}$). Both (i) the improvement of the radius of convergence to as large as $|g_{p}/\Delta_{p}| \sim 0.4$ upon including the quartic contribution, and (ii) the reliable prediction of dispersive shifts in tens of MHz from analytical theory, confirm how the generalized SWT method developed here provides a powerful framework to capture the effect of time-dependent strong interactions. 
\par
While the example discussed here involves a monochromatic pump, the sequential SWT method can accommodate ‘colored’ pumps, arbitrary forms of $g(t)$ via a decomposition into a Fourier series, and frequency-modulation of system frequencies [see appendix~\ref{app:qubitmod}]. A specific case, relevant to experimental situations, is inclusion of residual or parasitic static coupling present due to practical design limitations. The effect of such a static interaction can be captured by including a zero-frequency component in Eq.~(\ref{eqn:g_general})
\begin{equation}
g(t)= \sum\limits_{j=p, s}[g_j \exp(-i\omega_j t)+g_j^* \exp(+i\omega_j t)].
\end{equation}
Following the same procedure as outlined earlier, the time-averaged dispersive shift is the sum of different frequency components of the interaction, ${\chi_{p}^{(2)}= \sum_{j}\chi_{j}^{(2)}}$.
This provides a means to cancel out the net combined shift by canceling the shift caused by the static coupling against that caused by the parametric coupling. Note that this can be accomplished even when $|g_p| < |g_s|$, so long as a viable $\omega_p$ can be found. 
%
\subsection{Validity of rotating-wave approximation}
%
\begin{figure*}[t!]
  \begin{minipage}[b]{0.49\textwidth}
    \includegraphics[width=\textwidth]{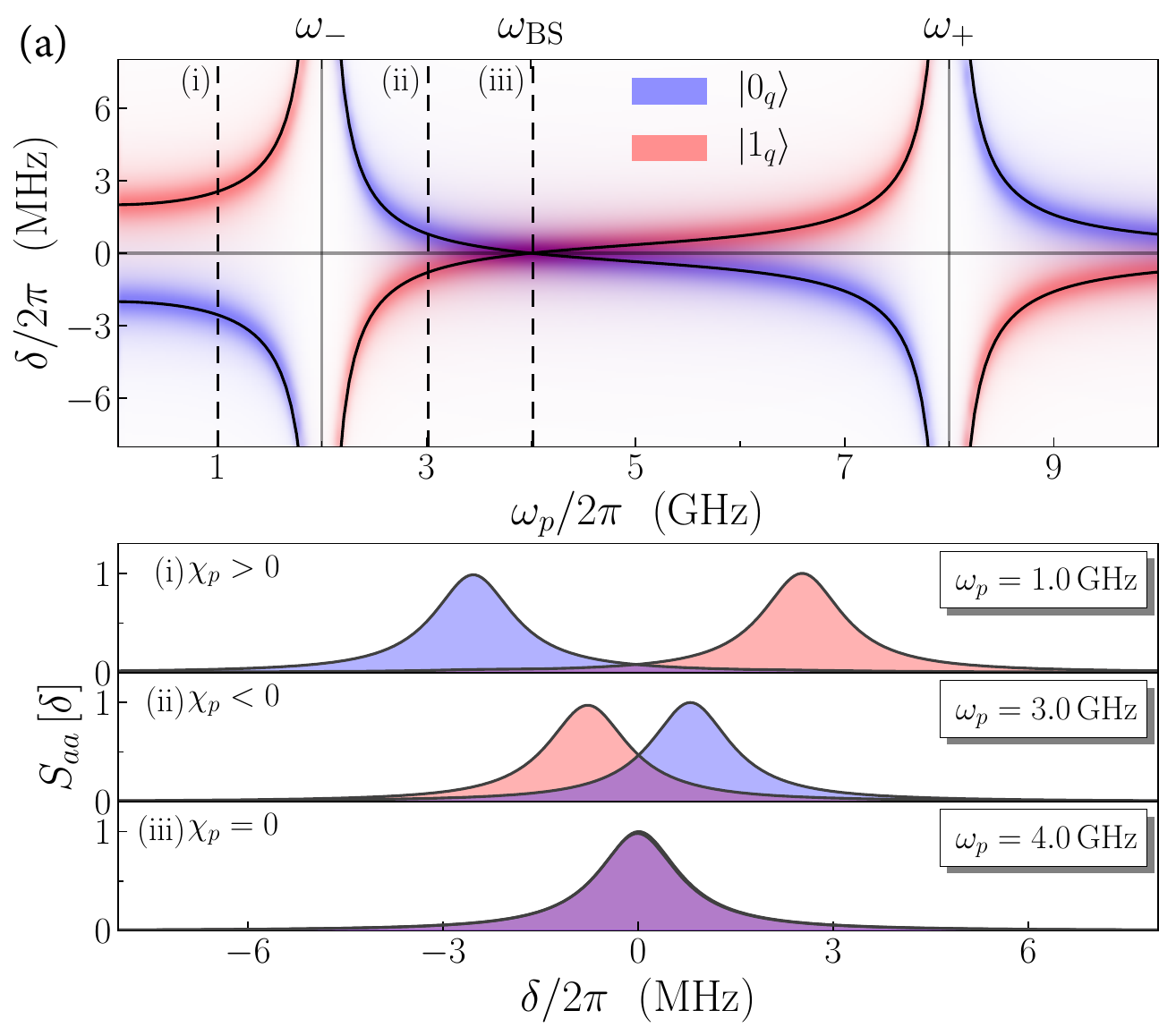}
  \end{minipage}
    \hfill
  \begin{minipage}[b]{0.49\textwidth}
    \includegraphics[width=\textwidth]{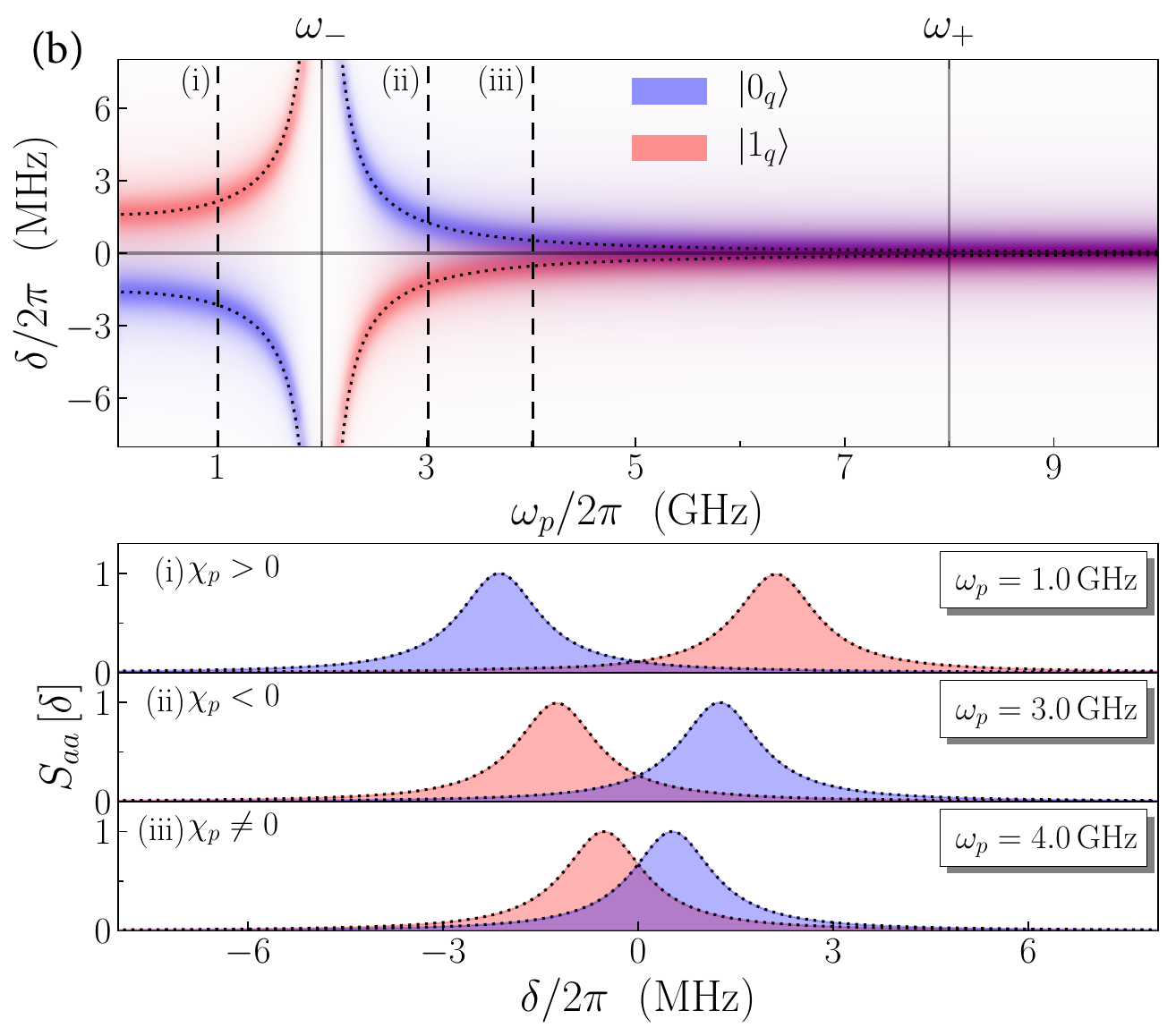}
  \end{minipage}
  	\caption{(a) Comparison of analytical results the resonator frequency shifts calculated using Eq.~\eqref{eqn:QubitDispersiveShift} (black) with the resonator spectrum $S_{aa}[\delta]$, for $\delta$ the detuning from $\omega_a$, obtained from exact numerical simulations (color) of the time-dependent Rabi Hamiltonian. The resonator spectral density is shown as a function of the pump frequency $\omega_p$ for qubit in the ground (blue) and excited (red) state. Bottom panel shows the normalized resonator spectra for selected values of $\omega_p$ demonstrating positive, negative, and zero $\chi_p$ (`blind spot') for Rabi model. (b) Analogous comparison for Jaynes-Cummings Hamiltonian. Note that no blind spot occurs in the absence of Bloch-Siegert contribution mediated by the sum frequency terms. All simulations were performed with $\omega_q=2\pi\times5$ GHz, $\omega_a=2\pi\times3$ GHz, $g_p=2\pi\times40$ MHz, and $\kappa=2\pi\times1.5$ MHz.}
\label{Fig:QR-JC-BlindSpot}
\vspace{-10pt}
\end{figure*}
A crucial consideration for strong interactions is the validity of the Hamiltonian derived under rotating-wave approximation (RWA). This is especially relevant for periodically-driven systems where, depending on the choice of $\omega_{p}$, any (or even a combination) of the time-dependent terms can be rendered resonant. Here we elucidate how it is imperative to make the RWA after deriving the transformed $H_{\rm eff}(t)$ to a given order, by contrasting the results obtained with the full Rabi interaction to the predictions for a parametrically-driven JC interaction. 
\par
To delineate the importance of the counter-rotating terms, $\hat{\sigma}_q^{+}\hat{a}^{\dagger} + {\rm h.c.}$, in the parametric-dispersive regime [Eq.~(\ref{eqn:validdisp})], we now derive the dressed shifts for a qubit-resonator system coupled via a parametrically-mediated Jaynes Cumminngs (JC) interaction of the form ${\hat{V}^{(1)}_{\text{JC}}(t)= g(t)(\hat{\sigma}_q^{+}\hat{a}+\hat{\sigma}_q^{-}\hat{a}^\dagger)}$. In this case, unlike Eq.~(\ref{eqn:SWT_Opertor}), the time-dependent SWT generator includes only the difference frequency components in line with the JC Hamiltonian,
\begin{equation}
\hat{S}^{(1)}_{\text{JC}}(t)=-{\xi}^{(1)}_{-}(t)\hat{\sigma}_q^{-}\hat{a}^\dagger-\text{h.c}
\end{equation}
The resultant leading-order effective Hamiltonian in the interaction picture is given by
\begin{equation}\begin{split}\label{eqn:H_eff_full_JC}
\hat{H}_\text{JC}^{(2)} (t)
=\chi_{\text{JC}}^{(2)}(t)\hat{\sigma}_q^z\left(\hat{a}^\dagger\hat{a}+\frac{1}{2}\right)+\Omega_{\text{JC}}^{(2)}(t),
\end{split}\end{equation}
where $\chi_{\text{JC}}^{(2)}(t)=-g(t)\text{Re}\left[{\xi}^{(1)}_{-}(t)\right]$ and the overall energy shift $\Omega_{\text{JC}}^{(2)}(t)=\frac{1}{2}g(t)\text{Re}\left[{\xi}^{(1)}_{-}(t)\right]$. Note that the qubit-state-dependent squeezing component of the effective Hamiltonian present in the Rabi model case does not appear for JC interaction. Considering a single-frequency pump as before, it leads to a simple expression for the time-averaged parametric disperive shift
\begin{equation}\begin{split}\label{eqn:JCdispshift}
\chi_{\text{JC}}^{(2)}=-2|g_p|^2\frac{\omega_{-}}{\omega_{-}^2-\omega_p^2}.
\end{split}\end{equation} 
As evident, the dispersive shift for JC Hamiltonian lacks the sum frequency contribution as a result of the missing counter-rotating terms in the Hamiltonian, contrasting the dispersive shift for Rabi Hamiltonian in Eq.~(\ref{eqn:QubitDispersiveShift}.
\par
Figure~\ref{Fig:QR-JC-BlindSpot} presents a detailed comparison of analytically calculated qubit-induced dispersive shifts on the resonator frequency, as a function of modulation frequency $\omega_{p}$, against the backdrop of exact resonator transmission spectrum obtained from a numerical simulation for both the Rabi and JC Hamiltonians. As evident, parametric interactions support a unique `blind spot' where $\chi_{p}=0$, even for large $g_{p}/\Delta_{p}$, for an appropriate pump frequency at the geometric average of the sum and the difference frequencies $\omega_{\text{BS}}=(\omega_{+}\omega_{-})^{1/2} = (\omega_q^2-\omega_a^2)^{1/2}$. (Note that no blind spot exists for $\omega_q<\omega_a$.) This decoupling is caused due to equal and opposite contributions to the dressed energy-level shifts from the difference and the sum frequency components of the Rabi Hamiltonian at $\omega_{\rm BS}$ respectively. Note that no blind spot exists for $\omega_q<\omega_a$. Such a cancellation at the two-level approximation is not possible with shifts estimated in Eq.~(\ref{eqn:JCdispshift}) with JC interaction, of which the profile of qubit-induced shifts on the resonator frequency for various values of $\omega_{p}$ against the backdrop of exact resonator transmission spectrum obtained from a numerical simulation is shown in Figure~\ref{Fig:QR-JC-BlindSpot}(b).
\par
In the same vein, the JC Hamiltonian fails to capture the pole near the sum frequency $\omega_{+}$ entirely. This is clear from Eq.~(\ref{eqn:JCdispshift}), which unlike Eq.~(\ref{eqn:QubitDispersiveShift}), lacks the sum frequency contribution. In fact, the contribution of the missing sum frequency terms can be significant even at pump frequencies far detuned from $\omega_{+}$; this is most strikingly evident at $\omega_{\rm BS}$, where a non-zero value of $\chi_{\text{JC}}$ persists due to the absence of Bloch-Siegert contributions. Thus, unlike the Rabi model, the JC interaction does not support zero-contrast points with monochromatic driving. Both these aspects highlight how making the rotating-wave approximation in the lab frame, before the SWT is performed, can be problematic and lead to erroneous results for time-dependent interactions. 
%
%

\section{Multi-level effects}
\label{sec:Kerr}
%
To investigate the effect of higher excitations in time-dependent settings, we next consider a multi-level generalization of the Rabi model by replacing the qubit with a Kerr oscillator. The Kerr oscillator provides an economical way to study the perturbative corrections for the case of a multi-frequency spectrum, while introducing only one additional parameter (anharmonicity $K$, with $K > 0$ assumed) to the problem. Further, it attracts considerable theoretical and practical interest since, besides being a canonical example of a nonlinear quantum optical system, it forms the cornerstone of many quantum information platforms using Josephson junction-based superconducting circuits \cite{Koch2007,Zhang2017,Andersen2020}.
\par
The system of a Kerr oscillator coupled to a linear oscillator via a time-dependent transverse coupling can be described by the Hamiltonian,
\begin{equation*}
\label{eqn:HKerr}
\hat{H}(t)=\omega_b \hat{b}^\dagger\hat{b}- K(\hat{b}^\dagger\hat{b})^2+\omega_a\hat{a}^\dagger\hat{a}+ g(t)(\hat{b}^\dagger+\hat{b})(\hat{a}^\dagger+\hat{a}),
\end{equation*}
where $\hat{b}^\dagger$~($\hat{a}^\dagger$), $\hat{b}$~($\hat{a}$) are creation and annihilation operators, $\omega_b$~($\omega_a$) is the frequency of the Kerr (linear) oscillator, and $g(t)=2g_p \cos(\omega_pt)$ as before. Following the same procedure as done for the qubit, we can define an SWT generator $S_{1}$ of the form
\begin{equation}
\begin{split}
\hat{S}^{(1)} (t)&=g_p \sum_{\pm} e^{\pm i \omega_p t} \left[\left(\hat{\Omega}_{+}[\hat{n}_{b}]\pm\omega_p\right)^{-1} \hat{b}^\dagger\hat{a}^\dagger\right. \\
& \hspace{2.5cm}+ \left.\left(\hat{\Omega}_{-}[\hat{n}_{b}]\pm\omega_p\right)^{-1}\hat{b}^\dagger\hat{a}\right]-\text{h.c.}, 
\end{split}
\end{equation}
where $\hat{\Omega}_{\pm}[\hat{n}_{b}] = (\omega_{b} + K)\hat{\mathbb{I}} - 2K \hat{n}_{b} \pm\omega_a \hat{\mathbb{I}}$ with $\hat{n}_{b}=\hat{b}^\dagger\hat{b}$. 
\begin{figure}[t!]
	\centering
    \includegraphics[width=0.98\columnwidth]{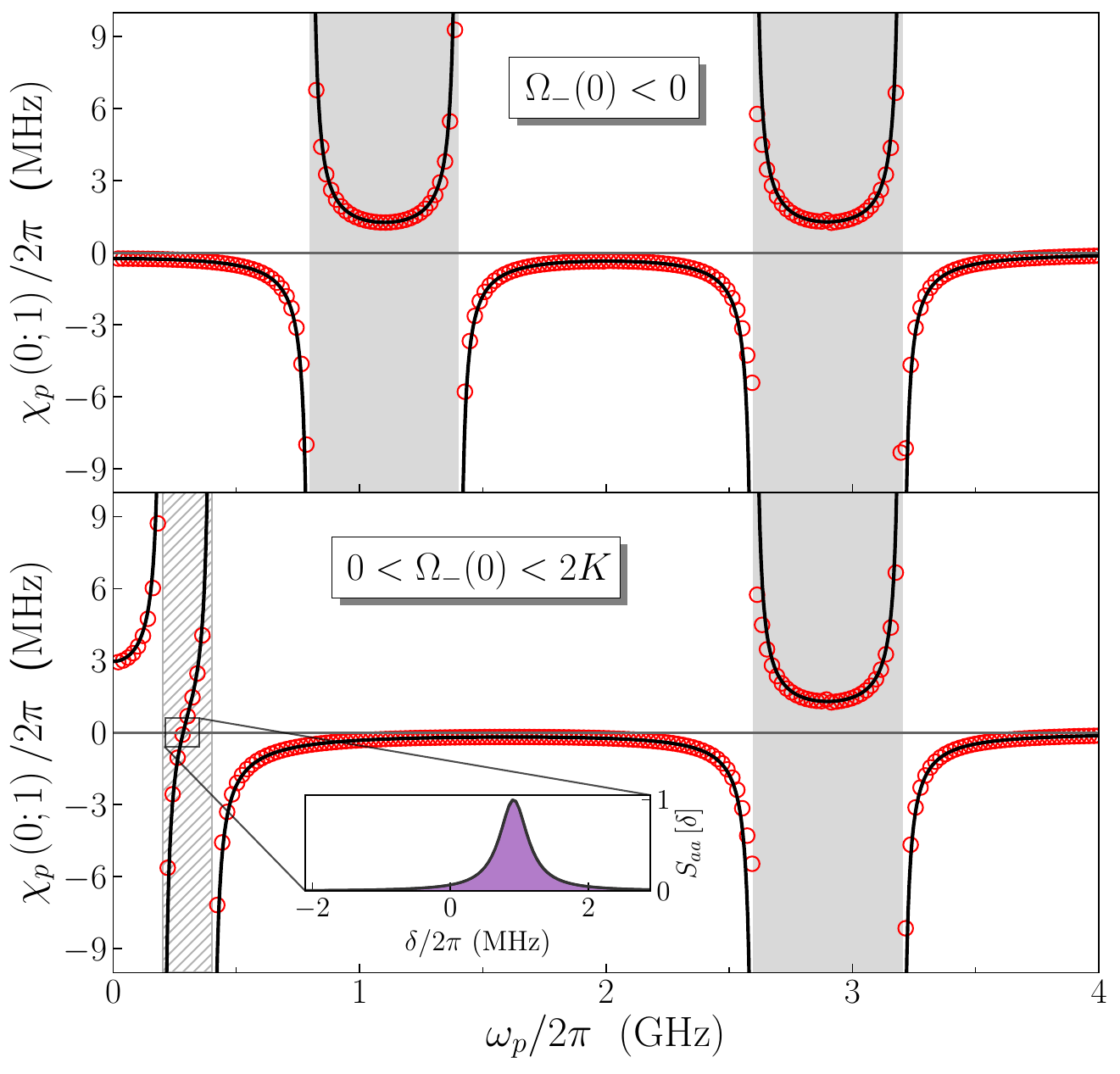}%
	\caption{
	Parametric dispersive shifts of a resonator coupled to a Kerr-oscillator for two distinct regimes of anharmonicity: $\omega_a=2\pi\times2.0\;$GHz, $\omega_b=2\pi\times1.5\;$GHz (top) and $\omega_a=2\pi\times1.5\;$GHz, $\omega_b=2\pi\times2.0\;$GHz (bottom). Both simulations use $K=2\pi\times300\;$MHz, $g_p=2\pi\times10\;$MHz, and $\kappa=2\pi\times0.5\;$MHz. The red circles are results from simulations of a truncated three-level system, and the black is the result of the analytical calculation based on  Eq.~\eqref{eqn:FirstOrderDispersiveShift}. In both panels, the gray regions correspond to the parametric straddling regimes, where $\chi_{p}(0;1)> 0$ is realized. The hatched area in the bottom panel corresponds to the region enclosing the blind spot. The inset shows the normalized resonator spectrum at the blind spot; note that while there is no state-dependent shift there is an overall shift of the resonator frequency at the $\omega_{\rm BS}$ unlike the qubit case [see appendix~\ref{app:gndstateshift}]. 
	}
	 \label{Fig:TR}%
\vspace{-5pt}%
\end{figure}%
\par
To leading order in $g_{p}$, we find the resultant cross-Kerr shift corresponding to the transition between the Fock states $\{|n_{b}-1\rangle, |n_{b}\rangle\}$ of the Kerr oscillator as
\begin{widetext}
\begin{equation}\label{eqn:FirstOrderDispersiveShift}
\chi_{p}^{(2)} (n_{b}-1; \;n_{b}) = g_p^2
\sum_{\pm}\left(\frac{n}{\Omega_{-}(n)\pm\omega_p} + \frac{n}{\Omega_{+}(n)\pm\omega_p}\left.-\frac{n+1}{\Omega_{-}(n+1)\pm\omega_p} -\frac{n+1}{\Omega_{+}(n+1)\pm\omega_p}\right)\right\vert^{n_{b}}_{n_{b}-1},
\end{equation}
\end{widetext}
where $\Omega_{\pm}(n_{b}) = \langle n_{b}|\hat{\Omega}_{\pm}[\hat{n}_{b}]|n_{b}\rangle$ denotes the number-dependent sum and difference frequencies for the system of coupled oscillators. Several observations are in order:
\begin{enumerate}[(i)]
\itemsep=0pt
\item The expression for the shift in Eq.~(\ref{eqn:FirstOrderDispersiveShift}) involves eight terms as compared to the two terms in Eq.~(\ref{eqn:QubitDispersiveShift}) for the two-level qubit case; this is because each state $|n_{b}\rangle$ is coupled to neighboring states $|n_{b} \pm 1\rangle$, each of which contribute to the net shift on the energy of a given state. 
\item Given the distinct number-dependent transition frequencies for the Kerr oscillator, the regime for validity of the perturbative diagonalization now depends on the number of excitations, i.e. $g_{p}/|\Delta_{p}(n_{b})| < 1$, where $\Delta_{p}(n_{b}) = {\rm min}\{\Omega_{\pm} (n_{b}) \pm \omega_{p} \}$.
\item We develop the perturbation series with $g_{p}/|\Delta_{p}(n_{b})|$ as the small parameter. This is a crucial improvement over some previous analyses, which treat the Kerr term perturbatively \cite{Nigg2012,Didier2018}, since now the shifts calculated using Eq.~(\ref{eqn:FirstOrderDispersiveShift}) hold for any value of anharmonicity $K$ relative to detuning from the parametric resonance $\Delta_{p}(n_{b})$.
\end{enumerate}
To elucidate the last point, Fig.~\ref{Fig:TR} shows the detailed profile of calculated $\chi_{p}^{(2)}(0;1)$ (the `transmon' limit \cite{Bourassa2012}) in two distinct regimes based on the magnitude of the $K$ relative to the detuning $\Omega_{-}(0) $; in both cases we assume weak anharmonicity s.t. $K<\omega_{a,b}$.   
\par
\vspace{5pt}
\underline{Case I}: $\Omega_{-}(0) < 0 \; {\rm or}\; 0 < 2K < \Omega_{-}(0)$. As shown in Fig.~\ref{Fig:TR}, either positive or negative shifts can be realized depending on whether the ratio $K/\Delta_{p}(n_{b})$ is larger or smaller than unity. In contrast, the sign of the static shift $\chi_{s}^{(2)}(0;1)$, $\omega_{p}=0$ in Eq.~(\ref{eqn:FirstOrderDispersiveShift}), is fixed by the sign of the anharmonicity $K$ alone. Specifically, $\chi_{p}^{(2)}(0;1)$ reverses sign when $\Omega_{-}(2)< \omega_{p} < \Omega_{-}(1)$ or $\Omega_{+}(2) < \omega_{p} < \Omega_{+}(1)$. This {\it parametric-straddling regime} \cite{Noh2021}, is reminiscent of the static straddling regime achieved with fixed couplings for $\omega_{b}- 3K < \omega_{a} < \omega_{b} - K$ \cite{Srinvisan2011}, except now the sign reversal can be achieved by tuning the frequency $\omega_{p}$ of the coupling while keeping its magnitude $g_{p}$ fixed.
\par
\vspace{5pt}
\underline{Case II}: $0 < \Omega_{-}(0) < 2K $. Here, in addition to the sign reversal of induced frequency shifts between parametric straddling and dispersive regimes, a blind spot $\chi_{p}^{(2)}(0;1) = 0$ is realized for a pump frequency $\omega_{\text{BS}}\simeq(\Omega_{-}(1)|\Omega_{-}(2)|)^{1/2}$. Note that, unlike the two-level qubit case, this does not result from the cancellation of Bloch-Siegert (sum) and JC (difference) contributions, but rather from the difference-frequency contributions to $\chi_{p}(0;1)$ corresponding to the lowest two transition frequencies of the Kerr oscillator. 
%
%
\section{Induced dissipation}
\label{sec:Purcell}
%
In the presence of parametric interactions, the dissipative interaction of the qubit with the environment (through the resonator) also becomes tunable with the pump frequency and amplitude, in sharp contrast with the static scenario where the induced dissipation is fixed \cite{Govia2015}. Using a formal master equation construction [see appendix~\ref{app:purcell}], we find that for $|\omega_{p}-\omega_{-}|\sim g_{p}$ the dominant process is qubit relaxation ($\hat{\sigma}_{q}^{-}$) at rate $\gamma_{-}^{\downarrow}$, while for $|\omega_{p}-\omega_{+}|\sim g_{p}$ induced qubit heating ($\hat{\sigma}_{q}^{+}$) at rate $\gamma_{+}^{\uparrow}$ dominates, with the respective rates given by
\begin{subequations}
\begin{align}
&  \gamma_{-}^{\downarrow} \approx \sum_{\pm}\kappa(\omega_{q}\pm \omega_{p}) \left(\frac{g_p}{\omega_{q} - \omega_{a}\pm\omega_{p}}\right)^2, \\ 
& \gamma_{+}^{\uparrow} \approx \sum_{\pm} \kappa(-\omega_{q}\pm \omega_{p})\left(\frac{g_p}{\omega_{q} + \omega_{a}\pm\omega_{p}}\right)^2.
\end{align}
\label{eq:Purcellrate}%
\end{subequations}
Here $\kappa(\omega)$ denotes the rate proportional to the noise spectral density of the environment at frequency $\omega$. Similar calculation for the case of the Kerr resonator leads to a $n_{b}$-dependent relaxation (heating) rate when the interaction is driven near the difference (sum) frequency, $\Omega_{\pm}(n_{b})$, for the corresponding number-dependent transition [see appendix~\ref{app:purcell}]. Note that the `quantum heating' rate $\gamma_{+}^{\uparrow}$ is non-zero even when the resonator remains in its ground state, since it is mediated through coupling-induced amplification of vacuum fluctuations \cite{Dykman2011}. Crucially, the induced dissipation rates do {\it not} null at the blind spots unlike the coherent shifts; they remain highly suppressed though since the corresponding pump frequency is far detuned from both the sum and difference frequencies, i.e. $|\omega_{\pm} \pm \omega_{\rm BS}|\gg g_{p}$. 
\par
As evident from Eq.~(\ref{eq:Purcellrate}), a fixed frequency qubit parametrically-coupled to an environment be employed as a novel broadband noise sensor. The resultant qubit relaxation or heating rates sample the environmental density of states over a set bandwidth (set by Lorentzian profile of resonator in QED setups), with the center of this `filter function' set by the choice of the pump frequency mediating the interaction. Such pump-mediated noise spectroscopy can be thought of as a continuous-wave analogue of spin-locking with the baseband frequency set by $\omega_{p}$ in GHz, making it immune to low-frequency fluctuations that typically limit conventional protocols relying on variation of the drive amplitude \cite{Slichter2012}. It is also an attractive complement to dynamical decoupling protocols which characterize slow or quasi-static noise \cite{Bylander2011}, and can be used for sensing high-frequency quantum noise which was shown to limit coherence in recent qubit designs \cite{Yan2016}.
%
%
\section{Conclusions}
\label{sec:conclusion}
%
We have presented an {\it ab-initio} framework for diagonalizing strong time-dependent interactions, based on a time-dependent generalization of the Schrieffer-Wolff transformation (SWT). The excellent quantitative agreement between exact numerical simulations and our analytical calculations for perturbative shifts and decay of dressed states demonstrates how the SWT can be used for developing a well-controlled perturbation series in time-dependent settings. The time-dependent SWT method presents a complementary approach to exact methods such as Floquet theory. Though exact, the complexity of Floquet-based diagonalization grows exponentially with the number of driving fields making it computationally demanding. On the other hand, while perturbative, SWT-based expansion is more amenable in terms of analytical tractability of the effective Hamiltonian thereby offering physically relevant insights. In fact, recent works have used SWT-like techniques for block diagonalization of the Floquet quasienergy operator in order to gain a more intuitive picture about validity of high-frequency approximations employed while truncating the Floquet Hilbert space \cite{Eckardt2015}.

Applying our analytical technique to parametric interactions, we predict several unique features accessible with parametric cavity- and circuit-QED systems. Most notably tunable dispersive shifts in tens of MHz can be realized with modest interaction strengths by simply tuning the pump frequency sufficiently close to the sum or difference frequency, while maintaining large detuning between the physical subsystems. This is a particularly attractive functionality for multi-qubit circuit-QED architectures where frequency crowding becomes a limiting issue \cite{Reagor2018}. Achieving such tunability in architectures based on static coupling and tunable qubits necessarily comes with trade-offs, such as flux noise-limited qubit coherence and enhanced crosstalk, leading to limited control flexibility. Further, our results show how the physics of the Rabi model can be crucial even in the dispersive regime $g_{p} \ll |\omega_{-}|$ and leads to significant corrections beyond the usual rotating-wave Jaynes-Cummings model; these are crucial for predicting exact cancellation points (`blind spots') of the induced energy shifts. We note that these corrections to the dressed shift are distinct from non-RWA corrections in the ultra-strong coupling ($g < \omega_{-} \lesssim \omega_{a}$) \cite{Zeuco2009} or the deep-strong coupling ($\omega_{-} < g \sim \omega_{a}$) \cite{Casanova2010} regime. In addition to their theoretical novelty, identifying and engineering such blind spots is a topic of active interest in several applications, such as protecting qubits against photon shot-noise induced dephasing \cite{Sears2012, ZhangHouck2017} and mitigating $ZZ$-induced crosstalk in two-qubit gates \cite{Mundada2019, Ku2020, Ganzhorn2020, Petrescu2021} in circuit-QED architectures. Extending our analysis to weakly anharmonic systems, we show how frequency-tunable interactions support a rich structure of multi-level shifts, including switching between parametric-dispersive and parametric-straddling regimes via the choice of pump frequency. Previously, such effects have been restricted to the purview of highly nonlinear multi-level atoms with complicated selection-rule engineering \cite{Zhu2013}. 

The parametric-QED regime presented here enables several new and expeditious applications of the burgeoning parametric quantum toolbox. In addition, they provide a natural starting point for several theoretical extensions, such as multi-pump generalizations, the multi-photon Rabi model \cite{Bertet2002}, and Hamiltonian engineering for quantum simulation of non-trivial gauge structures \cite{Goldman2014}, that we hope to investigate in future work. 
\begin{acknowledgments}
This research was supported by the Department of Energy under grant DE-SC0019461, with partial support from Air Force Office of Scientific Research under grant FA9550-21-1-0151.
\end{acknowledgments}
\appendix
%
%
\section{Time-dependent Schrieffer-Wolff Transformation (SWT)}
\label{app:SWT}
The essence of constructing a sequence of successive time-dependent SWTs that perturbatively diagonalizes a general time-dependent Hamiltonian is to find appropriate choice of $M$-th order SWT generator $\hat{S}^{(M)}(t)$ such that the off-diagonal part of the interaction Hamiltonian only contains terms of $\mathcal{O}(\lambda^M)$ or higher, i.e.
\begin{eqnarray}
\hat{V}_{\text{OD}}^{(M)}(t)=\sum\limits_{m=M}^{\infty}\lambda^{m}\hat{V}_{\text{OD},m}^{(M)}(t).  
\label{Eq:VODapp}
\end{eqnarray}
The form of this generator can be calculated using Eq.~(\ref{eqn:S_Generator}) presented in the main text; in this section we present a detailed proof of this construction. 
\par
We construct this proof by following a process similar to mathematical induction. Without loss of generality, we begin by assuming that the lab frame interaction Hamiltonian $\hat{V}(t)\sim \mathcal{O}(\lambda)$, such that $\hat{H}^{(1)}(t)$ satisfies Eq.~(\ref{Eq:VOD}) by construction. Then, assuming $\hat{H}^{(M)}(t)$ satisfies Eq.~(\ref{Eq:VOD}),  we will now prove that $\hat{H}^{(M+1)}(t)$ also satisfies Eq.~(\ref{Eq:VOD}) as long as $\hat{S}^{(M)}(t)$ satisfies Eq.~(\ref{eqn:S_Generator}). 
\par
Utilizing the Baker-Campbell-Hausdorff expansion and the derivative of the exponential map, the unitary transformation from $\hat{H}^{(M)}(t)$ to $\hat{H}^{(M+1)}(t)$ can be expressed as, with $\hat{U}_{M}(t)=\exp[-\hat{S}^{(M)}(t)]$,
%
\begin{widetext}
\begin{align}\begin{split}\label{eqn:SWT_total_Hamiltonian}
\hat{H}^{(M+1)}(t)
&=\hat{U}_{M}^{\dagger}(t)\hat{H}^{(M)}(t)\hat{U}_{M}(t)-i\hat{U}_{M}^{\dagger}(t)\frac{\partial\hat{U}_{M}(t)}{\partial t}\\
&=\hat{H}_0+\sum_{n=0}^{\infty}\frac{1}{(1+n)!}\left(\text{ad}_{\hat{S}^{(M)}(t)}\right)^{n}[\hat{S}^{(M)}(t),\hat{H}_0] +\sum_{n=0}^{\infty}\frac{1}{n!}\left(\text{ad}_{\hat{S}^{(M)}(t)}\right)^{n}\hat{V}^{(M)}(t)\\
&\hspace{0.9cm}+\sum_{n=0}^{\infty}\frac{1}{(1+n)!}\left(\text{ad}_{\hat{S}^{(M)}(t)}\right)^{n}\left(i\frac{\partial\hat{S}^{(M)}(t)}{\partial t}\right)\\
&=\hat{H}_0+\frac{1}{1!}\hat{W}^{(M)}(t)+\left(\frac{1}{1!}-\frac{1}{2!}\right)[\hat{S}^{(M)}(t),\hat{V}^{(M)}(t)]+\frac{1}{2!}[\hat{S}^{(M)}(t),\hat{W}^{(M)}(t)]\\
& \hspace{0.9cm} +\left(\frac{1}{2!}-\frac{1}{3!}\right)[\hat{S}^{(M)}(t),[\hat{S}^{(M)}(t),\hat{V}^{(M)}(t)]] +\frac{1}{3!}[\hat{S}^{(M)}(t),[\hat{S}^{(M)}(t),\hat{W}^{(M)}(t)]]\\
& \hspace{0.9cm}+\left(\frac{1}{3!}-\frac{1}{4!}\right)[\hat{S}^{(M)}(t),[\hat{S}^{(M)}(t),[\hat{S}^{(M)}(t),\hat{V}^{(M)}(t)]]] +\frac{1}{4!}[\hat{S}^{(M)}(t),[\hat{S}^{(M)}(t),[\hat{S}^{(M)}(t),\hat{W}^{(M)}(t)]]]+\cdots\\
&=\hat{H}_0 + \hat{W}^{(M)}(t) +  \sum_{n=1}^{\infty}\frac{1}{(n+1)!}\left(\text{ad}_{\hat{S}^{(M)}(t)}\right)^{n}\hat{W}^{(M)}(t) + \sum_{n=1}^{\infty}\frac{n}{(n+1)!}\left(\text{ad}_{\hat{S}^{(M)}(t)}\right)^{n}\hat{V}^{(M)}(t),
\end{split}
\end{align}
\end{widetext}%
%
%
\clearpage
\newpage
\mbox{~}
\clearpage
\newpage
\noindent where $\text{ad}_{\hat{S}^{(M)}(t)}\bullet\equiv[\hat{S}^{(M)}(t),\bullet]$ denotes the adjoint action, and
\begin{equation}\label{eqn:W}
\hat{W}^{(M)}(t)
=i\frac{\partial\hat{S}^{(M)}(t)}{\partial t}+[\hat{S}^{(M)}(t),\hat{H}_0]+\hat{V}^{(M)}(t).
\end{equation}
%
Next, following the line of inductive reasoning we assume that $\hat{H}^{(M)}$ satisfies Eq.~(\ref{Eq:VOD}). Therefore, the interaction Hamiltonian can be written as 
%

\begin{equation}\begin{split}
\hat{V}^{(M)}(t)&=\hat{V}_{\text{D}}^{(M)}(t)+\sum\limits_{m=M}^{\infty}\lambda^{m}\hat{V}_{\text{OD},m}^{(M)}(t)\\
&=\hat{V}^{(M)}_{\text{D}}(t)+\lambda^M\hat{V}_{\text{OD},M}^{(M)}(t)+\sum\limits_{m=M+1}^{\infty}\lambda^{m}\hat{V}_{\text{OD},m}^{(M)}(t),
\end{split}\end{equation}%
where we have singled out the lowest order off-diagonal term $\lambda^M\hat{V}_{\text{OD},M}^{(M)}(t)$ that we aim to eliminate. Given Eq.~(\ref{eqn:S_Generator}) for $\hat{S}^{(M)}(t)$, Eq.~(\ref{eqn:W}) reduces to
%
\begin{equation}\begin{split}\label{eqn:WM}
\hat{W}^{(M)}(t)
&=i\frac{\partial\hat{S}^{(M)}(t)}{\partial t}+[\hat{S}^{(M)}(t),\hat{H}_0]+\hat{V}^{(M)}_{D}(t)\\
&\quad+\lambda^M\hat{V}_{\text{OD},M}^{(M)}(t)+\sum\limits_{m=M+1}^{\infty}\lambda^{m}\hat{V}_{\text{OD},m}^{(M)}(t) \nonumber\\
&=\hat{V}^{(M)}(t)-\lambda^M\hat{V}_{\text{OD},M}^{(M)}(t)\nonumber\\
&=\hat{V}^{(M)}_{\text{D}}(t)+\sum\limits_{m=M+1}^{\infty}\lambda^{m}\hat{V}_{\text{OD},m}^{(M)}(t),
\end{split}\end{equation}
%
which is exactly the remainder of $\hat{V}^{(M)}$ without its lowest order off-diagonal component. Note that $\hat{V}^{(M)}(t), \hat{W}^{(M)}(t)$ are both at least of the order of $\mathcal{O}(\lambda^{n\geq1})$, while $\hat{S}^{(M)}(t) \sim \mathcal{O}(\lambda^{M})$ according to Eq.~(\ref{eqn:S_Generator}). Thus 
\begin{equation}\begin{split}\label{eqn:HigerOrderM+1}
&\sum_{n=1}^{\infty}\frac{1}{(n+1)!}\left(\text{ad}_{\hat{S}^{(M)}(t)}\right)^{n}\hat{W}^{(M)}(t)\sim \mathcal{O}(\lambda^{n\geq M+1}),\\
&\sum_{n=1}^{\infty}\frac{n}{(n+1)!}\left(\text{ad}_{\hat{S}^{(M)}(t)}\right)^{n}\hat{V}^{(M)}(t)
\sim \mathcal{O}(\lambda^{n\geq M+1})
\end{split}\end{equation}
Substituting Eq.~(\ref{eqn:WM}, \ref{eqn:HigerOrderM+1}) in Eq.~(\ref{eqn:SWT_total_Hamiltonian}), we arrive at the desired expression for $\hat{H}^{(M+1)}(t)$, 
\begin{equation}\begin{split}\label{eqn:}
\hat{H}^{(M+1)}(t)
=&\hat{H}_0 +\hat{V}^{(M)}_{\text{D}}(t)+\mathcal{O}(\lambda^{M+1}).
\end{split}\end{equation}
Note that $\hat{H}^{(M+1)}(t)$ lacks any $\mathcal{O}(\lambda^{M})$ order off-diagonal term. Thus, by constructing a time-dependent Schrieffer-Wolff generator using Eq.~(\ref{eqn:S_Generator}), $\hat{H}^{(M)}(t)$ can be transformed into $\hat{H}^{(M+1)}(t)$ with the off-diagonal component at $M^{\rm th}$ order eliminated. Since this property of the transformation is valid for any $M\geq1$, a sequence of successive SWTs can be applied to eliminate the off-diagonal terms up to any desired order in $\lambda$. 
\par
Note that although formally infinite number of terms are involved in the expression of $\hat{W}^{(M)}(t), \hat{V}^{(M)}(t), \hat{W}^{(M+1)}(t), \hat{V}^{(M+1)}(t)\cdots$, in practice none of these calculations need to be done beyond the desired order of diagonalization in $\lambda$, since they are anyway neglected in the final outcome. 
\vspace{15pt}
\subsection*{Example for $M_{\rm{max}}=5$}
%
As an illustrative example, we detail the procedure of obtaining a diagonalized Hamiltonian correct to 5-th order (i.e. $M_{\text{max}}=5$), obtained by eliminating off-diagonal interaction terms to the $\mathcal{O}(\lambda^4)$. To this end, without loss of generality, we begin by assuming $\hat{V}^{(1)}_{\text{D}}=0$ and write
\begin{equation}\begin{split}
\hat{V}^{(1)}(t)= \lambda\hat{V}^{(1)}_{\text{OD}}(t);
\end{split}\end{equation}
Using Eq.~(\ref{eqn:W}), this leads to the following differential equation for $\hat{S}^{(1)}(t)$,
\begin{equation}\begin{split}
i\frac{\partial\hat{S}^{(1)}(t)}{\partial t}+[\hat{S}^{(1)}(t),\hat{H}_0]+\lambda\hat{V}^{(1)}_{\text{OD}}(t)=0.
\end{split}\end{equation}
Therefore,
\begin{equation}\begin{split}
\hat{W}^{(1)}(t)=i\frac{\partial\hat{S}^{(1)}(t)}{\partial t}+[\hat{S}^{(1)}(t),\hat{H}_0]+\hat{V}^{(1)}(t)=0
\end{split}\end{equation}
According to Eq.~(\ref{eqn:SWT_total_Hamiltonian})
\begin{equation}\begin{split}
&\hat{V}^{(2)}(t)\\
=& \frac{1}{2}\left[\hat{S}^{(1)}(t),\hat{V}^{(1)}(t)\right]
+\frac{1}{3}\left[\hat{S}^{(1)}(t),\left[\hat{S}^{(1)}(t),\hat{V}^{(1)}(t)\right]\right]\\
&+\frac{1}{8}\left[\hat{S}^{(1)}(t),\left[\hat{S}^{(1)}(t),\left[\hat{S}^{(1)}(t),\hat{V}^{(1)}(t)\right]\right]\right]
+\mathcal{O}(\lambda^{5}).
\end{split}\end{equation}
%
Next, following this procedure the second-order generator $\hat{S}^{(2)}(t)$ can be evaluated from
\begin{equation}\begin{split}
i\frac{\partial\hat{S}^{(2)}(t)}{\partial t}+[\hat{S}^{(2)}(t),\hat{H}_0]+\lambda^2\hat{V}^{(2)}_{\text{OD},2}(t)=0.
\end{split}\end{equation}
where $\hat{V}^{(2)}_{\text{OD},2}(t)$ is entirely generated from the off-diagonal contribution due to action of the previous SWT on the interaction, i.e., 
\begin{eqnarray}
\lambda^2\hat{V}^{(2)}_{\text{OD},2}(t)=\mathcal{Q}_{0} \bullet\frac{1}{2}\left[\hat{S}^{(1)}(t),\hat{V}^{(1)}(t)\right].
\label{Eq:VOD2}
\end{eqnarray}
Substituting this in the expression for $\hat{W}^{(2)}$, we obtain
%
\begin{widetext}
\begin{equation}\begin{split}
\hat{W}^{(2)}&=i\frac{\partial\hat{S}^{(2)}(t)}{\partial t}+[\hat{S}^{(2)}(t),\hat{H}_0]+\hat{V}^{(2)}(t) \\
&=i\frac{\partial\hat{S}^{(2)}(t)}{\partial t}+[\hat{S}^{(2)}(t),\hat{H}_0]+\mathcal{Q}_{0}\bullet\frac{1}{2}\left[\hat{S}^{(1)}(t),\hat{V}^{(1)}(t)\right]+\mathcal{P}_{0}\bullet\frac{1}{2}\left[\hat{S}^{(1)}(t),\hat{V}^{(1)}(t)\right]\\
&\quad +\frac{1}{3}\left[\hat{S}^{(1)}(t),\left[\hat{S}^{(1)}(t),\hat{V}^{(1)}(t)\right]\right]
+\frac{1}{8}\left[\hat{S}^{(1)}(t),\left[\hat{S}^{(1)},\left[\hat{S}^{(1)}(t),\hat{V}^{(1)}(t)\right]\right]\right]
+\mathcal{O}(\lambda^{5})\\
&=\mathcal{P}_{0}\bullet\frac{1}{2}\left[\hat{S}^{(1)}(t),\hat{V}^{(1)}(t)\right]
+\frac{1}{3}\left[\hat{S}^{(1)}(t),\left[\hat{S}^{(1)}(t),\hat{V}^{(1)}(t)\right]\right]
+\frac{1}{8}\left[\hat{S}^{(1)}(t),\left[\hat{S}^{(1)}(t),\left[\hat{S}^{(1)}(t),\hat{V}^{(1)}(t)\right]\right]\right]
+\mathcal{O}(\lambda^{5})
\end{split}\end{equation}
Next, according to Eq.~(\ref{eqn:SWT_total_Hamiltonian}),
\begin{equation}\begin{split}
\hat{V}^{(3)}(t) &= \hat{W}^{(2)}(t)  +\frac{1}{2}\left[\hat{S}^{(2)}(t),\hat{W}^{(2)}(t)\right]+\frac{1}{6}\left[\hat{S}^{(2)}(t),\left[\hat{S}^{(2)}(t),\hat{W}^{(2)}(t)\right]\right] \nonumber\\
& \hspace{1.6cm} +\frac{1}{2}\left[\hat{S}^{(2)}(t),\hat{V}^{(2)}(t)\right]+\frac{1}{3}\left[\hat{S}^{(2)}(t),\left[\hat{S}^{(2)}(t),\hat{V}^{(2)}(t)\right]\right]+\mathcal{O}(\lambda^{8})\\
&\simeq\hat{W}^{(2)}(t)  +\frac{1}{2}\left[\hat{S}^{(2)}(t),\hat{W}^{(2)}(t)\right]
+\frac{1}{2}\left[\hat{S}^{(2)}(t),\hat{V}^{(2)}(t)\right]+\mathcal{O}(\lambda^{5})\\
&=\mathcal{P}_{0}\bullet\frac{1}{2}\left[\hat{S}^{(1)}(t),\hat{V}^{(1)}(t)\right]
+\frac{1}{3}\left[\hat{S}^{(1)}(t),\left[\hat{S}^{(1)}(t),\hat{V}^{(1)}(t)\right]\right]
+\frac{1}{8}\left[\hat{S}^{(1)}(t),\left[\hat{S}^{(1)}(t),\left[\hat{S}^{(1)}(t),\hat{V}^{(1)}(t)\right]\right]\right]\\
& \quad+\frac{1}{2}\left[\hat{S}^{(2)}(t),\mathcal{P}_{0}\bullet\frac{1}{2}\left[\hat{S}^{(1)}(t),\hat{V}^{(1)}(t)\right]\right]
+\frac{1}{2}\left[\hat{S}^{(2)}(t),\hat{V}^{(2)}(t)\right]+\mathcal{O}(\lambda^{5}),
\end{split}\end{equation}
where in the second step we drop the terms $\left[\hat{S}^{(2)}(t),\left[\hat{S}^{(2)}(t),\hat{W}^{(2)}(t)\right]\right], \left[\hat{S}^{(2)}(t),\left[\hat{S}^{(2)}(t),\hat{V}^{(2)}(t)\right]\right]\sim\mathcal{O}(\lambda^{6})$, since we are only interested in elimination of off-diagonal terms up to $\mathcal{O}(\lambda^4)$. This leads to the following differential equation for third-order generator $\hat{S}^{(3)}(t)$
\begin{equation}\begin{split}
i\frac{\partial\hat{S}^{(3)}(t)}{\partial t}+[\hat{S}^{(3)}(t),\hat{H}_0]+\lambda^3\hat{V}^{(3)}_{\text{OD},3}(t)=0.
\end{split}\end{equation}
where now the only off-diagonal terms at cubic order are obtained as
\begin{equation}
\lambda^3\hat{V}^{(3)}_{\text{OD},3}(t)=\mathcal{Q}_{0}\bullet\frac{1}{3}\left[\hat{S}^{(1)}(t),\left[\hat{S}^{(1)}(t),\hat{V}^{(1)}(t)\right]\right]. 
\end{equation}
This leads to
\begin{equation}\begin{split}
\hat{W}^{(3)}&=i\frac{\partial\hat{S}^{(3)}(t)}{\partial t}+[\hat{S}^{(3)}(t),\hat{H}_0]+\hat{V}^{(3)}(t) \\
&=i\frac{\partial\hat{S}^{(3)}(t)}{\partial t}+[\hat{S}^{(3)}(t),\hat{H}_0]+\mathcal{P}_{0}\bullet\frac{1}{2}\left[\hat{S}^{(1)}(t),\hat{V}^{(1)}(t)\right]\nonumber\\
&\quad+\mathcal{Q}_{0}\bullet\frac{1}{3}\left[\hat{S}^{(1)}(t),\left[\hat{S}^{(1)}(t),\hat{V}^{(1)}(t)\right]\right]+\mathcal{P}_{0}\bullet\frac{1}{3}\left[\hat{S}^{(1)}(t),\left[\hat{S}^{(1)}(t),\hat{V}^{(1)}(t)\right]\right]\\
&\quad+\frac{1}{8}\left[\hat{S}^{(1)}(t),\left[\hat{S}^{(1)}(t),\left[\hat{S}^{(1)}(t),\hat{V}^{(1)}(t)\right]\right]\right]+\frac{1}{2}\left[\hat{S}^{(2)}(t),\mathcal{P}_{0}\bullet\frac{1}{2}\left[\hat{S}^{(1)}(t),\hat{V}^{(1)}(t)\right]\right]
+\frac{1}{2}\left[\hat{S}^{(2)}(t),\hat{V}^{(2)}(t)\right]
+\mathcal{O}(\lambda^{5})\\
&=\mathcal{P}_{0}\bullet\frac{1}{2}\left[\hat{S}^{(1)}(t),\hat{V}^{(1)}(t)\right]
+\mathcal{P}_{0}\bullet\frac{1}{3}\left[\hat{S}^{(1)}(t),\left[\hat{S}^{(1)}(t),\hat{V}^{(1)}(t)\right]\right]
+\frac{1}{8}\left[\hat{S}^{(1)}(t),\left[\hat{S}^{(1)}(t),\left[\hat{S}^{(1)}(t),\hat{V}^{(1)}(t)\right]\right]\right]\\
&\quad+\frac{1}{2}\left[\hat{S}^{(2)}(t),\mathcal{P}_{0}\bullet\frac{1}{2}\left[\hat{S}^{(1)}(t),\hat{V}^{(1)}(t)\right]\right]
+\frac{1}{2}\left[\hat{S}^{(2)}(t),\hat{V}^{(2)}(t)\right]+\mathcal{O}(\lambda^{5}).
\end{split}\end{equation}
Noting that $\hat{V}^{(3)}(t),\hat{W}^{(3)}(t)\sim\mathcal{O}(\lambda^{2})$ and $\hat{S}^{(3)}(t)\sim\mathcal{O}(\lambda^{3})$, we can ignore terms of the type $\left[\hat{S}^{(3)}(t),\hat{W}^{(3)}(t)\right]$, $\left[\hat{S}^{(3)}(t),\hat{V}^{(3)}(t)\right]$ which are $\mathcal{O}(\lambda^{5})$, since we are interested in obtaining the diagonalized Hamiltonian to $\mathcal{O}(\lambda^{4})$. Therefore,
\begin{equation}\begin{split}
\hat{V}^{(4)}(t) &= \hat{W}^{(3)}(t)  +\frac{1}{2}\left[\hat{S}^{(3)}(t),\hat{W}^{(3)}(t)\right]+ \mathcal{O}(\lambda^{9})
+\frac{1}{2}\left[\hat{S}^{(3)}(t),\hat{V}^{(3)}(t)\right]+ \mathcal{O}(\lambda^{8})\\
&\simeq\hat{W}^{(3)}(t) + \mathcal{O}(\lambda^{5})\\
&= \mathcal{P}_{0}\bullet\frac{1}{2}\left[\hat{S}^{(1)}(t),\hat{V}^{(1)}(t)\right]
+\mathcal{P}_{0}\bullet\frac{1}{3}\left[\hat{S}^{(1)}(t),\left[\hat{S}^{(1)}(t),\hat{V}^{(1)}(t)\right]\right]
+\frac{1}{8}\left[\hat{S}^{(1)}(t),\left[\hat{S}^{(1)}(t),\left[\hat{S}^{(1)}(t),\hat{V}^{(1)}\right]\right]\right]\\
& \quad +\frac{1}{2}\left[\hat{S}^{(2)}(t),\mathcal{P}_{0}\bullet\frac{1}{2}\left[\hat{S}^{(1)}(t),\hat{V}^{(1)}(t)\right]\right]
+\frac{1}{2}\left[\hat{S}^{(2)}(t),\hat{V}^{(2)}(t)\right]+\mathcal{O}(\lambda^{5})
\end{split}\end{equation}
Next, the differential equation for $\hat{S}^{(4)}(t)$ becomes
\begin{equation}\begin{split}
i\frac{\partial\hat{S}^{(4)}(t)}{\partial t}+[\hat{S}^{(4)}(t),\hat{H}_0]+\lambda^4\hat{V}^{(4)}_{\text{OD},4}(t)=0.
\end{split}\end{equation}
where the three off-diagonal contributions at $\mathcal{O}(\lambda^4)$ order are given as

\begin{equation}\begin{split}
\lambda^4\hat{V}^{(4)}_{\text{OD},4}(t)=&\mathcal{Q}_{0}\bullet\left(\frac{1}{8}\left[\hat{S}^{(1)}(t),\left[\hat{S}^{(1)}(t),\left[\hat{S}^{(1)}(t),\hat{V}^{(1)}(t)\right]\right]\right]
+\frac{1}{2}\left[\hat{S}^{(2)}(t),\mathcal{P}_{0}\bullet\frac{1}{2}\left[\hat{S}^{(1)}(t),\hat{V}^{(1)}(t)\right]\right]
+\frac{1}{2}\left[\hat{S}^{(2)}(t),\hat{V}^{(2)}(t)\right]\right).
\end{split}\end{equation}
Therefore,
\begin{equation}\begin{split}
\hat{W}^{(4)}(t)&=\mathcal{P}_{0}\bullet\frac{1}{2}\left[\hat{S}^{(1)}(t),\hat{V}^{(1)}(t)\right]
+\mathcal{P}_{0}\bullet\frac{1}{3}\left[\hat{S}^{(1)}(t),\left[\hat{S}^{(1)}(t),\hat{V}^{(1)}(t)\right]\right]
+\mathcal{P}_{0}\bullet\frac{1}{8}\left[\hat{S}^{(1)}(t),\left[\hat{S}^{(1)}(t),\left[\hat{S}^{(1)}(t),\hat{V}^{(1)}(t)\right]\right]\right]\\
&\quad+\mathcal{P}_{0}\bullet\left(\frac{1}{2}\left[\hat{S}^{(2)}(t),\mathcal{P}_{0}\bullet\frac{1}{2}\left[\hat{S}^{(1)}(t),\hat{V}^{(1)}(t)\right]\right]\right)
+\mathcal{P}_{0}\bullet\frac{1}{2}\left[\hat{S}^{(2)}(t),\hat{V}^{(2)}(t)\right]+\mathcal{O}(\lambda^{5})
\end{split}\end{equation}
and
\begin{equation}\begin{split}
\hat{V}^{(5)}(t)&= \hat{W}^{(4)}(t) + \frac{1}{2}\left[\hat{S}^{(4)}(t),\hat{W}^{(4)}(t)\right]+\frac{1}{2}\left[\hat{S}^{(4)}(t),\hat{V}^{(4)}(t)\right]+\mathcal{O}(\lambda^{10})\\
&\simeq\hat{W}^{(4)}(t) +\mathcal{O}(\lambda^{6})\\
&=\mathcal{P}_{0}\bullet\frac{1}{2}\left[\hat{S}^{(1)}(t),\hat{V}^{(1)}(t)\right]
+\mathcal{P}_{0}\bullet\frac{1}{3}\left[\hat{S}^{(1)}(t),\left[\hat{S}^{(1)}(t),\hat{V}^{(1)}(t)\right]\right]
+\mathcal{P}_{0}\bullet\frac{1}{8}\left[\hat{S}^{(1)}(t),\left[\hat{S}^{(1)}(t),\left[\hat{S}^{(1)}(t),\hat{V}^{(1)}(t)\right]\right]\right]\\
&\quad+\mathcal{P}_{0}\bullet\left(\frac{1}{2}\left[\hat{S}^{(2)}(t),\mathcal{P}_{0}\bullet\frac{1}{2}\left[\hat{S}^{(1)}(t),\hat{V}^{(1)}(t)\right]\right]\right)
+\mathcal{P}_{0}\bullet\frac{1}{2}\left[\hat{S}^{(2)}(t),\hat{V}^{(2)}(t)\right]+\mathcal{O}(\lambda^{5})
\end{split}\end{equation}
In the examples discussed in the main text, $\left[\hat{S}^{(1)}(t),\left[\hat{S}^{(1)}(t),\hat{V}^{(1)}(t)\right]\right]$ and $\left[\hat{S}^{(2)}(t),\mathcal{P}_{0}\bullet\left[\hat{S}^{(1)}(t),\hat{V}^{(1)}(t)\right]\right]$ generate no diagonal contributions, therefore,
\begin{equation}\begin{split}
\hat{V}^{(5)}(t)
=\mathcal{P}_{0}\bullet\left(\frac{1}{2}\left[\hat{S}^{(1)}(t),\hat{V}^{(1)}(t)\right]
+\frac{1}{8}\left[\hat{S}^{(1)}(t),\left[\hat{S}^{(1)}(t),\left[\hat{S}^{(1)}(t),\hat{V}^{(1)}(t)\right]\right]\right]
+\frac{1}{2}\left[\hat{S}^{(2)}(t),\hat{V}^{(2)}(t)\right]\right)+\mathcal{O}(\lambda^{5}).\qquad
\end{split}\end{equation}
\end{widetext}
%
%
\section{Time-dependent system frequency}
\label{app:qubitmod}
%
%
Our framework can fully accommodate the case where qubit frequency is time-dependent due to, say parametric flux variation. To demonstrate this, we take the Rabi model as an example and calculate only up to $\hat{S}^{(1)}(t)$. The total Hamiltonian now becomes $\hat{H}(t)=\hat{H}_0^{\text{TD}}(t)+\hat{V}(t)$,
where
%
$\hat{H}_0^{\text{TD}}(t)=-(\omega_q(t)/2)\hat{\sigma}_q^z+\omega_a\left(\hat{a}^\dagger\hat{a}+1/2\right).$
%
It is more convenient to obtain the SWT generator by first transforming into the interaction picture, where
\begin{equation}\begin{split}
\hat{V}_{\text{I}}(t)
=\exp{\left[i\int_{0}^{t} \hat{H}_0^{\text{TD}}(t') dt'\right]}\hat{V}(t)\exp{\left[-i\int_{0}^{t} \hat{H}_0^{\text{TD}}(t') dt'\right]},
\end{split}\end{equation}
and $i\partial\hat{S}_{\text{I}}^{(1)}/\partial t +\hat{V}_{\text{I}}(t)=0$.
We can then solve for
\begin{equation}
\hat{S}_{\text{I}}^{(1)}(t)=\tilde{{\xi}}_{+}(t)\hat{\sigma}_q^{+}\hat{a}^\dagger-\tilde{{\xi}}_{-}(t)\hat{\sigma}_q^{-}\hat{a}^\dagger - \text{h.c.},
\end{equation}
where 
\begin{equation}
\tilde{{\xi}}_{\pm}(t)=\pm i \int_{0}^{t} g(t') \exp{\left\{i \int_{0}^{t'} \left[\pm\omega_q(t'')+\omega_a\right] dt''\right\}}dt'.
\label{eq:SWTwithrect1}
\end{equation}
\par
As a concrete example, we consider the frequently encountered case of parametric flux variation of the qubit frequency, $\Phi(t) = \Phi_{0}+ \delta\Phi\cos(\omega_{d} t)$ with $|\delta\Phi| \ll \Phi_{0}$. Such nonlinear modulation of the qubit frequency can be written as
\begin{eqnarray*}
\omega_{q}[\Phi(t)] = \omega_{q}\cos[\Phi_{0}]+ \delta\Phi\cos(\omega_{d} t)] \approx \omega_{q0} + \nu\cos(\omega_{d} t),
\end{eqnarray*}
where $\omega_{q0} = \omega_{q}\cos[\Phi_{0}]$ and $\nu = \delta\Phi \times \omega_{q}\sin[\Phi_{0}]$. Substituting this in Eq.~(\ref{eq:SWTwithrect1}) and making use of Jacobi–Anger expansion to resolve the exponent of the sinusoidal modulation, we obtain
%
\begin{widetext}
\begin{equation}\begin{split}
\tilde{{\xi}}_{\pm}(t)&=\pm i \int_{0}^{t} g(t') \exp{\left\{i \left[\left(\pm\omega_{q0}+\omega_a\right) t' \pm \frac{\nu}{\omega_d}\sin{(\omega_d t')}\right]\right\}}dt'\\
&=\pm \sum_{k=-\infty}^{+\infty} J_k(\nu/\omega_d) \sum_{p} \left\{\frac{g_p}{\pm\omega_{q0} \pm k \omega_d+\omega_a-\omega_p} \exp{\left[ i \left(\pm\omega_{q0} \pm k \omega_d+\omega_a-\omega_p\right) t\right]}\right.\\
&\hspace{5cm}\left.+\frac{g_p^*}{\pm\omega_{q0} \pm k \omega_d+\omega_a+\omega_p} \exp{\left[ i \left(\pm\omega_{q0} \pm k \omega_d+\omega_a+\omega_p\right) t\right]}\right\},
\end{split}\end{equation}
where we have assumed sinusoidal coupling as the main text Eq.~(\ref{eqn:g_general}). Here $J_k(\nu/\omega_d)$ is the Bessel function of of the first kind, and $k$ is an integer. The expression for the corresponding SWT generator in the Schr\"{o}dinger picture,
\begin{equation}\begin{split}
\hat{S}^{(1)}(t)
=\tilde{{\xi}}_{+}(t)\exp{\left\{-i \int_{0}^{t} \left[+\omega_q(t')+\omega_a\right] dt'\right\}}\hat{\sigma}_q^{+}\hat{a}^\dagger
-\tilde{{\xi}}_{-}(t)\exp{\left\{-i \int_{0}^{t} \left[-\omega_q(t')+\omega_a\right] dt'\right\}}\hat{\sigma}_q^{-}\hat{a}^\dagger - \text{h.c.}
\label{eq:SWTwithrect2}
\end{split}\end{equation}
leads to
\begin{equation}\begin{split}
{\xi}_{\pm}(t)
&=\pm \sum_{k, k'=-\infty}^{+\infty} J_k(\nu/\omega_d) J_{k'}(\nu/\omega_d) \sum_{p} \left\{\frac{g_p}{\pm\omega_{q0} \pm k \omega_d+\omega_a-\omega_p} e^{ i \left[\pm (k-k') \omega_d -\omega_p\right] t}\right.\\
&\hspace{6cm}\left.+\frac{g_p^*}{\pm\omega_{q0} \pm k \omega_d+\omega_a+\omega_p} e^{ i \left[\pm (k-k') \omega_d +\omega_p\right] t}\right\}.
\end{split}\end{equation}
This can now be used in $\hat{H}^{(2)}(t) =\frac{1}{2}[\hat{S}^{(1)}(t),\hat{V}^{(1)}(t)] +\mathcal{O}(g_{p}^{3})$ [Eq.~(\ref{eqn:H_eff_full})] to calculate the leading-order effective Hamiltonian  correct to $\mathcal{O}(g^{2}(t))$. For a single pump frequency $\omega_p$, the corresponding dispersive shift, when $r \equiv 2\omega_p/\omega_d$ is an integer, is
\begin{equation}\begin{split}\label{}
\chi_{p}^{(2)} &=-|g_p|^2 \sum_{k=-\infty}^{+\infty} J_k^2(\nu/\omega_d) \left(\frac{1}{{{\omega_{k,-}}-\omega_p}}+\frac{1}{{{\omega_{k,-}}+\omega_p}}+\frac{1}{{{\omega_{k,+}}-\omega_p}}+\frac{1}{{{\omega_{k,+}}+\omega_p}}\right)\\
& \quad-\frac{1}{2} (g_p^2+{{g_p}^*}^2) \sum_{k=-\infty}^{+\infty} \left(\frac{J_k(\nu/\omega_d)J_{k-r}(\nu/\omega_d) }{{{\omega_{k,-}}-\omega_p}}+\frac{J_k(\nu/\omega_d)J_{k+r}(\nu/\omega_d)}{{{\omega_{k,-}}+\omega_p}}\right.\\
&\hspace{5cm}\left.+\frac{J_k(\nu/\omega_d)J_{k-r}(\nu/\omega_d)}{{{\omega_{k,+}}-\omega_p}}+\frac{J_k(\nu/\omega_d)J_{k+r}(\nu/\omega_d)}{{{\omega_{k,+}}+\omega_p}}\right),
\end{split}\end{equation}
where $\omega_{\pm,k} = (\omega_{q0}+k\omega_d) \pm \omega_{a}$. If $r$ is not an integer,
\begin{equation}\begin{split}\label{}
\chi_{p}^{(2)} =- |g_p|^2 \sum_{k=-\infty}^{+\infty} J_k^2(\nu/\omega_d)\left(\frac{1}{{{\omega_{k,-}}-\omega_p}}
+\frac{1}{{{\omega_{k,-}}+\omega_p}}+\frac{1}{{{\omega_{k,+}}-\omega_p}}+\frac{1}{{{\omega_{k,+}}+\omega_p}}\right).
\end{split}\end{equation}
Note that now in addition to qubit frequency $\omega_{q0}$  we get contributions to dressed shift at every order from sidebands at $\omega_{q0}+k\omega_d$, weighted by $J_k(\nu/\omega_d)$ .
\end{widetext}
%
\clearpage
\newpage
\mbox{~}
\clearpage
\newpage
%
\section{Multi-level shifts}
\label{app:gndstateshift}
%
The leading order diagonalized Hamiltonian for the system of a Kerr oscillator parametrically-coupled to a linear oscillator, is given by,
\begin{eqnarray}\label{eqn:TRDispersiveShift}
\hat{H}_{D}^{(2)}&=& |g_p|^2 \hat{n}_{a} \sum_{\pm, \pm}^{}\bigg[\bigg(\hat{\Omega}_{\pm} [\hat{n}_{b}]\pm\omega_p\bigg)^{-1}\hat{n}_{b} \nonumber\\
& & \qquad \qquad \qquad-\bigg(\hat{\Omega}_{\pm}[\hat{n}_{b}+1]\pm\omega_p\bigg)^{-1}(\hat{n}_{b}+1)\bigg], \nonumber\\
\end{eqnarray}
where $\hat{\Omega}_{\pm}[\hat{n}_{b}] = (\omega_{b} + K)\hat{\mathbb{I}} - 2K \hat{n}_{b} \pm\omega_a \hat{\mathbb{I}}$. 
\par
\begin{figure}[t!]
\centering
\includegraphics[width=0.45\textwidth]{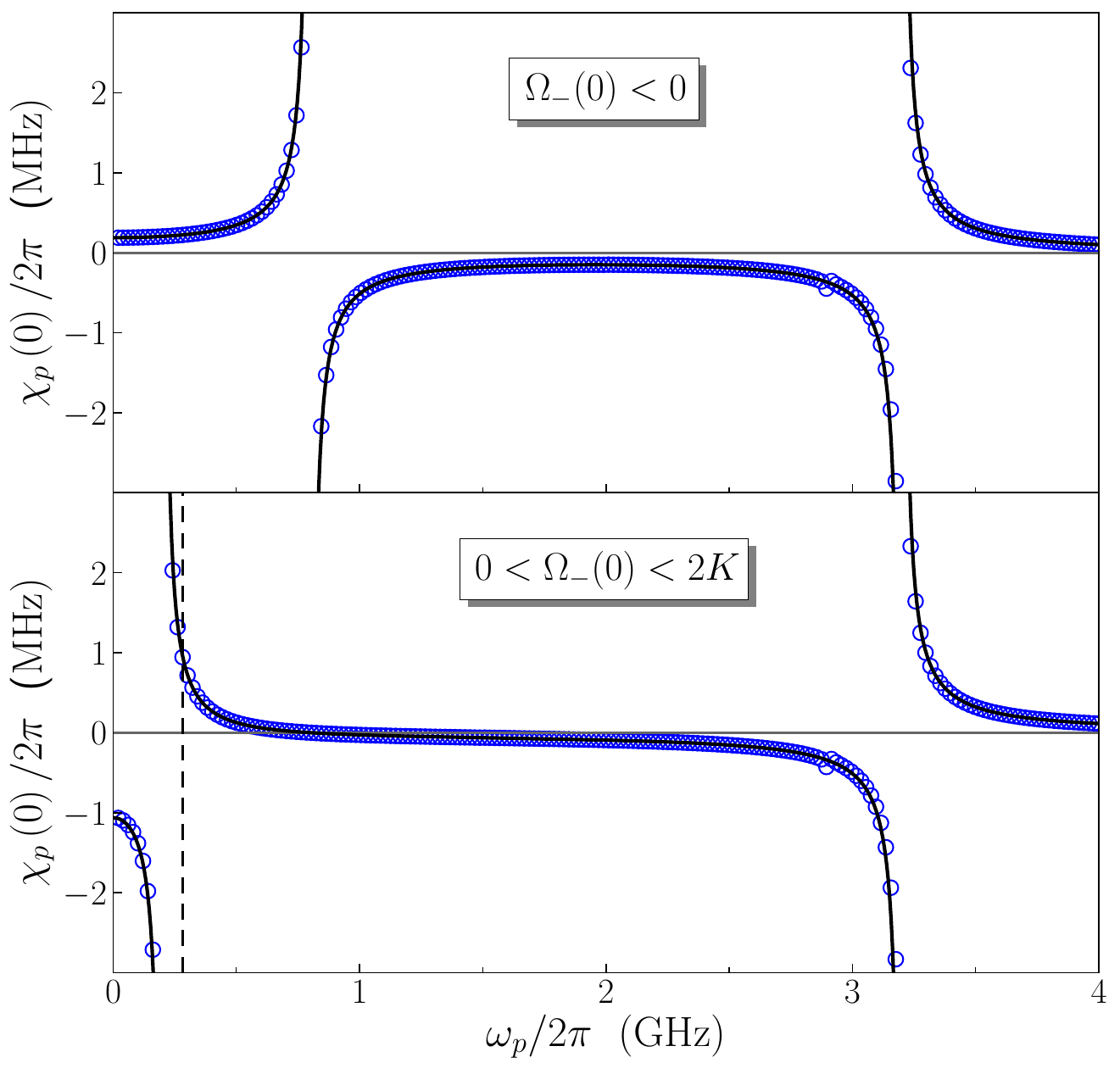}
 \caption{
    (Top) Shift of a resonator coupled to a three-level system in its ground state, with $\omega_a=2\pi\times1.5\;$GHz, $\omega_b=2\pi\times2.0\;$GHz, $K=2\pi\times300\;$MHz, $g_p=2\pi\times10\;$MHz, and $\kappa=2\pi\times0.5\;$MHz. The blue circles are results from simulations, and the black line is from Eq.~\ref{eqn:GroundStateDispersiveShift}
	(Bottom) Shift with $\omega_a=2\pi\times2.0\;$GHz, $\omega_b=2\pi\times1.5\;$GHz, and all other parameters as before. 
}
\label{Fig:QRBlindSpot}
\end{figure}

In addition to cross-Kerr shift reported in the maint text, the dispersive shift induced by the Kerr oscillator ``ground-state" can be calculated from the matrix element 
\begin{equation}\begin{split}\label{eqn:GroundStateDispersiveShift}
\chi_{p}^{(2)}(0) = \langle n_{a}, n_{b}=0|\hat{H}_{D}^{(2)}|n_{a}, n_{b}=0\rangle.
\end{split}\end{equation}
Fig.~\ref{Fig:QRBlindSpot} shows a profile of $\chi_{p}(0)$ for the two regimes of nonlinearity discussed in the main text. It is interesting to note that even at the `blind spot' realized in the regime $0 < \Omega_{-}(0) < 2K $, $\chi_{p}(0) \neq 0$ [see the inset of Fig.~4 in the main text]; this is in contrast to the qubit case where the cancellation happens at zero detuning from the bare resonance of the linear oscillator.
\vspace{15pt}
\section{Higher-order corrections}
\label{app:higherorder}
%
The higher-order contributions in $\lambda$ can be developed using the prescription given in appendix~\ref{app:SWT}. Here we report the next order contribution in the systems we consider in the main text, which appears at quartic order in the coupling $g_p^4$ for the purely block off-diagonal interactions considered. Notably, there are two contributions at quartic order: (i) the diagonal term generated by the action of the first-order generator on the first-order off-diagonal interaction $\frac{1}{8}[\hat{S}^{(1)}(t),[\hat{S}^{(1)}(t),[\hat{S}^{(1)}(t),\hat{V}(t)]]]$, and (ii) the diagonal term generated by the second-order generator on the second-order off-diagonal interaction, $\frac{1}{2}[\hat{S}^{(2)}(t),\hat{V}_{\rm OD}^{(2)}(t)]$ [see Eq.~(\ref{Eq:VOD2})]. It is worth noting that the contribution of type (ii) does not appear for the JC interaction, as the source of the non-zero $\hat{V}_{\rm OD}^{(2)}(t)$ are the squeezing terms. Here we report the terms generated for both these interactions post-RWA for both the qubit and the Kerr resonator cases.
%
\subsection{Qubit case}
%
The two terms that appear in the diagonalized Hamitlonian at quartic order are
\begin{widetext}
\begin{eqnarray}
& & \frac{1}{8}[\hat{S}^{(1)}(t),[\hat{S}^{(1)}(t),[\hat{S}^{(1)}(t),\hat{V}(t)]]] \nonumber\\
& & \quad=
|g_p|^4\left[\sum\limits_{\pm}^{}\left(\frac{1}{\omega_{\pm}+\omega_p}+\frac{1}{\omega_{\pm}-\omega_p}\right)^3-\sum\limits_{\pm}^{}\left(\frac{1}{\omega_{\pm}+\omega_p}\right)^2\left(\frac{1}{\omega_{\pm}-\omega_p}\right)-\sum\limits_{\pm}^{}\left(\frac{1}{\omega_{\pm}-\omega_p}\right)^2\left(\frac{1}{\omega_{\pm}+\omega_p}\right)\right.\nonumber\\
& & \qquad\left.+2\sum\limits_{\pm}^{}\left(\frac{1}{\omega_{\pm}+\omega_p}+\frac{1}{\omega_{\pm}-\omega_p}\right)\left(\frac{1}{\omega_{\mp}+\omega_p}+\frac{1}{\omega_{\mp}-\omega_p}\right)^2\right.\nonumber\\
& & \qquad\left.-2\sum\limits_{\pm}^{}\left(\frac{1}{\omega_{\pm}+\omega_p}\right)\left(\frac{1}{\omega_{\mp}+\omega_p}\right)\left(\frac{1}{\omega_{\mp}-\omega_p}\right)\right. \nonumber\\
& & \qquad\left.-2\sum\limits_{\pm}^{}\left(\frac{1}{\omega_{\pm}-\omega_p}\right)\left(\frac{1}{\omega_{\mp}-\omega_p}\right)\left(\frac{1}{\omega_{\mp}+\omega_p}\right)\right]\hat{\sigma}_q^z\left[(\hat{a}^\dagger\hat{a})^2+\hat{a}^\dagger\hat{a}+\frac{1}{2}\right]
\end{eqnarray} 
and
\begin{eqnarray}
& & \frac{1}{2}[\hat{S}^{(2)}(t),\hat{V}_{\rm OD}^{(2)}(t)]=
-|g_p|^4
\left[ \frac{1}{\omega_a}\left(\sum_{\pm,\pm}\frac{1}{\omega_b\pm\omega_a\pm\omega_p}\right)^2+\frac{1}{\omega_a+\omega_p}\left(\frac{1}{\omega_b+\omega_a+\omega_p}+\frac{1}{\omega_b-\omega_a-\omega_p}\right)^2\right.
\nonumber\\
& & \hspace{4.5cm}\left.+\frac{1}{\omega_a-\omega_p}\left(\frac{1}{\omega_b-\omega_a+\omega_p}+\frac{1}{\omega_b+\omega_a-\omega_p}\right)^2\right]\frac{2\hat{a}^\dagger\hat{a}-1}{4}.
\end{eqnarray}
\end{widetext}
%
%
%
\subsection{Kerr resonator case}
%
For the Kerr resonator case, recall that $\chi_{p}^{(2)} (0;1)$, which is quadratic in $g_{p}$, exhibits poles at detunings of $-K$ and $-3K$ from the difference frequency $\omega_{-}$ [Fig.~3 of the main text]; these correspond to the `single-photon' parametric resonances for the $|0\rangle \rightarrow |1\rangle$ and $|1\rangle\rightarrow|2\rangle$ transitions of the Kerr resonator respectively. The next order corrections, quartic in $g_{p}$, describe the `two-photon' parametric resonances corresponding to the $|1\rangle \rightarrow |3\rangle$ transition of the Kerr resonator, thus manifesting as poles in $\chi_{p}^{(4)}(0;1)$ at $-5K$ and $+K$ detunings from the difference frequency. The resultant dispersive shifts can be read off from the prefactor of the $\hat{n}_{a} \hat{n}_{b}$ term in the SWT Hamiltonian diagonalized to fourth order, as
\begin{widetext}
\begin{equation}\begin{split}\label{eqn:ThirdOrderDispersiveShift}
&\chi_{p}^{(4)}(0;1)\\
=&-\frac{1}{4}|g_p|^4 \sum_{\pm}^{} \left[-18\left(\frac{1}{\Omega_{-}(3)\pm \omega_{p}}\right)\left(\frac{1}{\Omega_{-}(3)\pm \omega_{p}}\right)\left(\frac{1}{\Omega_{-}(2)\pm \omega_{p}}\right)\right.\\
&+ 36\left(\frac{1}{\Omega_{-}(2)\pm \omega_{p}}\right)\left(\frac{1}{\Omega_{-}(3)\pm \omega_{p}}\right)\left(\frac{1}{\Omega_{+}(2)\pm \omega_{p}}\right) +54\left(\frac{1}{\Omega_{+}(2)\pm \omega_{p}}\right)\left(\frac{1}{\Omega_{-}(3)\pm \omega_{p}}\right)\left(\frac{1}{\Omega_{+}(2)\pm \omega_{p}}\right)\\
&+ \left.6\left(\frac{1}{\Omega_{-}(0)\pm \omega_{p}}\right)\left(\frac{1}{\Omega_{-}(2)\pm \omega_{p}}\right)\left(\frac{1}{\Omega_{+}(1)\pm \omega_{p}}\right)+4\left(\frac{1}{\Omega_{+}(0)\pm \omega_{p}}\right)\left(\frac{1}{\Omega_{-}(1)\pm \omega_{p}}\right)\left(\frac{1}{\Omega_{+}(1)\pm \omega_{p}}\right)\right],\\
\end{split}\end{equation}
\end{widetext}
where we have considered the pump to be near the difference frequency. Note that the dressed shifts to cubic order in $g_{p}$ are zero under RWA, since diagonal contributions to shifts require a balanced number of $\hat{b}^\dagger$ and $\hat{b}$. 
\par
Crucially, the degree of each pole in the expression for the dispersive shift is connected to the order of the $n$-photon process. For instance, for $\omega_{p} = \Omega_{-}(3)$, the pole near $5K$ manifests as a peak in $\chi_{p}^{(4)}(0;1)$ which is symmetric in detuning, since it corresponds to a pole of degree 2 [first line of Eq.~(\ref{eqn:ThirdOrderDispersiveShift})]. This can be distinguished from the quadratic contribution to the dispersive shift $\chi_{p}^{(2)}(1;2)$ due to the $|2\rangle\rightarrow|3\rangle$ transition, which leads to a pole at the same value of $\omega_{p}$, but with a peak lineshape which is asymmetric in detuning as it is a pole of degree 1. 
%
\section{Parametrically-Induced Purcell decay and quantum heating}
\label{app:purcell}
%
To investigate the induced dissipation on the qubit or Kerr oscillator in the presence of a time-dependent interaction with a dissipative linear oscillator, in this section we construct a Lindblad-form quantum master equation. To this end, we transform the system-bath interaction using the same generator as that used to implement the time-dependent Schrieffer-Wolff transformation on the system Hamiltonian. Since this master equation construction is based on a second-order Dyson series expansion of the system-bath interaction, we restrict ourselves to the lowest-order operator transformations $\mathcal{O}(g_{p}/\Delta_{p}(n_{b}))$. 
\par
We present the results for the Kerr oscillator here since the qubit-case is a specific limit of the multi-level physics of this system. Specifically, our focus is obtaining the rate of induced, or Purcell, decay on a given transition of the Kerr oscillator, as well as the onset of dissipative terms mediated by the parametric coupling such as quantum heating. To this end, we assume the linear resonator is coupled to an environment modelled as a collection of harmonic oscillators. $\hat{H}_{E} = \sum_{\alpha} \nu_{\alpha} \hat{\gamma}_{\alpha}^{\dagger}\hat{\gamma}_{\alpha}$, with an interaction of the form
\begin{eqnarray}
\hat{H}_{SE}=\sum\limits_{\alpha}\mu_\alpha(\hat{\gamma}^\dagger_\alpha+\hat{\gamma}_\alpha)(\hat{a}^\dagger+\hat{a})
\end{eqnarray}
Under the action of the SWT, $\hat{H}_{SE}$ in the interaction frame transforms as,
\begin{widetext}
\begin{equation}\begin{split}\label{eqn:H_ES_SWT}
\hat{H}_{SE}^{\text{SW}}(t)
&= e^{\hat{S}(t)}\hat{H}_{SE}e^{-\hat{S}(t)}\\
&\simeq\sum_{\alpha}\mu_\alpha\left(\hat{\gamma}^\dagger_\alpha e^{i \nu_\alpha t}+\hat{\gamma}_\alpha e^{-i \nu_\alpha t}\right)\times \left[e^{-i \omega_at} 
\left\{\hat{a}- {g_p}\left({\hat{\Omega}_{+-}}^{-1}e^{i \hat{\Omega}_{+-} t} \hat{b}^\dagger
-\hat{b}\,{\hat{\Omega}_{-+}}^{-1}e^{-i \hat{\Omega}_{-+} t} \right)\right.\right.\\
&\left.\left.\hspace{6.5cm}-{g_p^*}\left({\hat{\Omega}_{++}}^{-1}e^{i \hat{\Omega}_{++} t}\hat{b}^\dagger-\hat{b}\,{\hat{\Omega}_{--}}^{-1}e^{-i \hat{\Omega}_{--} t}\right)\right\} + {\rm h.c.}\right],
\end{split}\end{equation}
\end{widetext}
where we have introduced the operator $\hat{\Omega}_{\pm\pm}[\hat{n}_{b}]= \hat{\Omega}_{\pm}[\hat{n}_{b}]\pm\omega_p\mathbb{I}$, with $\Omega_{\pm\pm} (n_{b}) \equiv \langle n_{b}|\hat{\Omega}_{\pm\pm}|n_{b}\rangle$, for brevity of notation. 
Note that if $g_p/|\Omega_{\pm\pm} (n_{b})|\rangle \ll 1$, the corresponding terms in Eq.~(\ref{eqn:H_ES_SWT}) are highly suppressed.
Taking the zero-temperature limit and considering a continuous density of states $\Gamma(\nu)$ for the bath modes leads to the following form of the master equation,
\begin{widetext}
\begin{equation}\begin{split}\label{eqn:MEfull}
\frac{\partial}{\partial t}\hat{\rho}_{I}(t)
&=(\hat{a}\hat{\rho}_{I}(t)\hat{a}^\dagger-\hat{a}^\dagger\hat{a}\hat{\rho}_{I}(t))\int\limits_{0}^{\infty}d\nu\int\limits_{0}^{\infty}dt'\;\Gamma(\nu)|\mu|^2 e^{i\nu(t'-t)} e^{-i\omega_a (t'-t)}+{\rm h.c.}\\
& \quad +(\hat{a}^\dagger\hat{\rho}_{I}(t)\hat{a}-\hat{a}\hat{a}^\dagger\hat{\rho}_{I}(t))\int\limits_{0}^{\infty}d\nu\int\limits_{0}^{\infty}dt'\;\Gamma(\nu)|\mu|^2 e^{i\nu(t'-t)} e^{+i\omega_a (t'-t)}+{\rm h.c.}\\
&\quad+(\hat{b}\hat{\rho}_{I}(t)\hat{b}^\dagger-\hat{b}^\dagger\hat{b}\hat{\rho}_{I}(t))\sum_{\pm}^{}\frac{g_p^2}{|\Omega_{-\pm} (n_{b}+1)|^2}\int\limits_{0}^{\infty}d\nu\int\limits_{0}^{\infty}dt'\;\Gamma(\nu)|\mu|^2 e^{i\nu(t'-t)} e^{-i(\omega_a + \Omega_{-\pm}(n_{b}+1)) (t'-t)}+{\rm h.c.}\\
&\quad+(\hat{b}^\dagger\hat{\rho}_{I}(t)\hat{b}-\hat{b}\hat{b}^\dagger\hat{\rho}_{I}(t))\sum_{\pm}^{}\frac{g_p^2}{|\Omega_{-\pm}(n_{b})|^2}\int\limits_{0}^{\infty}d\nu\int\limits_{0}^{\infty}dt'\;\Gamma(\nu)|\mu|^2 e^{i\nu(t'-t)} e^{+i(\omega_a + \Omega_{-\pm} (n_{b}))(t'-t)}+{\rm h.c.}\\
&\quad+(\hat{b}\hat{\rho}_{I}(t)\hat{b}^\dagger-\hat{b}^\dagger\hat{b}\hat{\rho}_{I}(t))\sum_{\pm}^{}\frac{g_p^2}{|\Omega_{+\pm} (n_{b}+1)|^2}\int\limits_{0}^{\infty}d\nu\int\limits_{0}^{\infty}dt'\;\Gamma(\nu)|\mu|^2 e^{i\nu(t'-t)} e^{+i(\omega_a - \Omega_{+\pm} (n_{b}+1)) (t'-t)}+{\rm h.c.}\\
&\quad+(\hat{b}^\dagger\hat{\rho}_{I}(t)\hat{b}-\hat{b}\hat{b}^\dagger\hat{\rho}_{I}(t))\sum_{\pm}^{}\frac{g_p^2}{|\Omega_{+\pm} (n_{b})|^2}\int\limits_{0}^{\infty}d\nu\int\limits_{0}^{\infty}dt'\;\Gamma(\nu)|\mu|^2 e^{i\nu(t'-t)} e^{-i(\omega_a -\Omega_{+\pm} (n_{b})) (t'-t)}+{\rm h.c.}
\end{split}\end{equation}
\end{widetext}
Before proceeding, it is worth clarifying the shorthand notation of Eq.~\eqref{eqn:MEfull}. As an example, we explicitly write out the following term in Eq.~\eqref{eqn:MEfull}
\begin{equation}\begin{split}
    &\hat{b}\hat{\rho}_I(t)\hat{b}^\dagger \frac{g_p^2}{|\Omega_{-+}(n_b)|^2}\\ 
    & \,= \sum_{n_b}\frac{n_b g_p^2}{|\Omega_{-+}(n_b)|^2}|n_{b}-1\rangle\langle n_{b}|
\hat{\rho}_I(t)|n_{b}\rangle\langle n_{b}-1|, \label{eqn:MEshorthand}
\end{split}\end{equation}
which shows that there are distinct disspiative rates for each Fock-state transition $n_b \rightarrow n_b -1$. To obtain the ``diagonal'' form of Eq.~\eqref{eqn:MEshorthand} we have made a RWA to drop terms of the form $|n_{b}-1\rangle\langle n_{b}|\hat{\rho}_I(t)|m_{b}\rangle\langle m_{b}-1|$ for $n_b \neq m_b$ during the master equation derivation. These terms would have a time-dependent phase factor $\exp\{-i\left[\Omega_{-+}(n_b) - \Omega_{-+}(m_b)\right]t\}$ in the interaction frame, which for $\Omega_{-+}(n_b) - \Omega_{-+}(m_b) = 2K(m_b-n_b)$ is fast oscillating. Note that we have a shifted transition frequency corresponding to $(n_{b}+1)$ excitations in the Kerr oscillator for $\hat{b}\bullet \hat{b}^{\dagger}$ due to non-commutation of $\hat{\Omega}_{\pm\pm}$ with $(\hat{b}, \hat{b}^{\dagger})$, namely,
\begin{equation}\begin{split}
\hat{b}^{\dagger}\hat{\Omega}_{\pm \pm} = (\hat{\Omega}_{\pm \pm} + 2K\, \mathbb{I})\hat{b}^{\dagger}, \quad
 \hat{b}\hat{\Omega}_{\pm \pm} = (\hat{\Omega}_{\pm \pm} - 2K\, \mathbb{I})\hat{b}.
\end{split}\end{equation}
Finally, the master equation can be written in a more compact form as
\begin{equation}\begin{split}
\frac{\partial}{\partial t}\hat{\rho}_{I}(t)
&= \bigg[\kappa(\omega_{a}) \mathcal{D}[\hat{a}]\\
&\quad+ \sum_{n_b}\left( \left[\gamma_{-}^{\downarrow}(n_b) + \gamma_{+}^{\downarrow}(n_b)\right] \mathcal{D}\left[|n_{b}-1\rangle\langle n_{b}|\right]\right.\\
&\quad\left.+ \left[\gamma_{-}^{\uparrow}(n_b) + \gamma_{+}^{\uparrow}(n_b)\right]\mathcal{D}\left[|n_{b}\rangle\langle n_{b}-1|\right]\right)\bigg]\hat{\rho}_{I}(t)
\end{split}\end{equation}
where $\mathcal{D}[\hat{O}]\bullet=\hat{O}\bullet\hat{O}^\dagger-\frac{1}{2}\hat{O}^\dagger\hat{O}\bullet-\frac{1}{2}\bullet\hat{O}^\dagger\hat{O}$, and $\kappa(\omega) = 2 \pi \Gamma(\omega) |\mu (\omega)|^2$, with $\gamma^\downarrow$ and $\gamma^\uparrow$ representing relaxation-type and heating-type dissipators respectively. The Kerr oscillator dissipative rates are given by
\begin{equation}\begin{split}
    &\gamma_{-}^{\downarrow}(n_b) = \sum_{\pm}^{}\kappa[\omega_{a} + \Omega_{-\pm}(n_{b})]\frac{g_p^2}{|\Omega_{-\pm}(n_{b})|^2},\\
    &\gamma_{-}^{\uparrow}(n_b) = \sum_{\pm}^{}\kappa[-\omega_{a} - \Omega_{-\pm}(n_{b})] \frac{g_p^2}{|\Omega_{-\pm}(n_{b})|^2},\\
    &\gamma_{+}^{\downarrow}(n_b) = \sum_{\pm}^{}\kappa[-\omega_{a} + \Omega_{+\pm}(n_{b})] \frac{g_p^2}{|\Omega_{+\pm}(n_{b})|^2},\\
    &\gamma_{+}^{\uparrow}(n_b) = \sum_{\pm}^{}\kappa[+\omega_{a} - \Omega_{+\pm}(n_{b})] \frac{g_p^2}{|\Omega_{+\pm}(n_{b})|^2}.
\end{split}\end{equation}
Typically, under the assumption that the bath modes remain in vacuum only the relaxation process is present; however, for parametric pumping, heating is possible even with a zero-temperature environment due to amplification of the zero-point
fluctuations.
Note that if the pump detuning from either the sum or difference frequency is large compared to $g_p$, i.e.~${\Omega_{\pm\pm}(n_{b})\gg g_p}$, then the corresponding dissipative rates are highly suppressed due to a large denominator in the respective prefactor. We focus on two distinct cases, depending on choice of pump frequency to be near the sum or difference frequency of the transmon-resonator system:
\vspace{5pt}
\par
\noindent \underline{Case I}: $\omega_{p}\approx |\omega_{-}|$\\
In this limit, depending on the detuning of the pump frequency, either $\Omega_{--}(n_{b})$ or $\Omega_{-+}(n_{b})$ is comparable in magnitude to  $g_p$, such that the corresponding rate dominates. Further, $\Omega_{-\pm}(n_{b})\ll \omega_a$, such that the corresponding $\kappa(-\omega_{a} - \langle\hat{\Omega}_{-\pm}\rangle)\sim\kappa(-\omega_{a}) =0$, since the bath spectrum has no negative frequency components. The result of this is that the prefactor for the corresponding heating term, $\mathcal{D}[\hat{b}^\dagger]$, is zero. Thus, whichever decay term is made non-negligible by the pump will have its counterpart heating term exactly zero. 
On the other hand, for whichever of $\Omega_{-\pm}(n_{b})$ is not comparable to $g_p$, we have that $\Omega_{-\pm} \gg g_p$, such that the corresponding decay term is highly suppressed, while its counterpart heating term can be potentially non-zero but nonetheless still highly suppressed. For both the sum frequency components we have that $\Omega_{+\pm}(n_{b})\gg g_p$, and thus, both the corresponding decay and heating rates are negligible. In summary, the dominant contribution to the dissipation comes from the parametrically-induced relaxation rate, which leads to the master equation
\begin{equation}\begin{split}\label{eqn:ME1}
\frac{\partial}{\partial t}\hat{\rho}_{I}(t)
\simeq\left[\kappa(\omega_{a}) \mathcal{D}[\hat{a}]+  \sum_{n_b}\gamma_{-}^{\downarrow}(n_b)\mathcal{D}[|n_{b}-1\rangle\langle n_{b}|]\right]\hat{\rho}_{I}(t).
\end{split}\end{equation} 
\par
\noindent \underline{Case II}: $\omega_{p}\approx \omega_{+}$\\
Following a similar analysis to the preceding discussion, in this regime parametrically-induced quantum heating dominates the dissipative dynamics, with the master equation taking the form
\begin{equation}\begin{split}\label{eqn:ME2}
\frac{\partial}{\partial t}\hat{\rho}_{I}(t)
\simeq\left[\kappa(\omega_{a}) \mathcal{D}[\hat{a}]+\sum_{n_b}\gamma_{+}^{\uparrow}(n_b)\mathcal{D}[|n_{b}\rangle\langle n_{b}-1|]\right]\hat{\rho}_{I}(t).
\\
\end{split}\end{equation} 
\par
As a final point, we emphasize that in either the qubit or the Kerr oscillator case, if the pump frequency is at the blind-spot frequency, the system still experiences induced decay and heating, even though the effective coherent coupling is exactly cancelled. Nonetheless, since the blind-spot frequency, $\omega_p = \omega_{BS}$, is far detuned from either the sum or difference frequency, both Purcell decay and heating effects, though non-zero, are highly suppressed due to $|g_{p}/\Omega_{p}| \ll 1$ where $\Omega_{p} = {\rm min}\{\Omega_{\pm\pm}(n_{b})\}$.
%
%
%

%
\end{document}